# QueryVis: Logic-based diagrams help users understand complicated SQL queries faster

Aristotelis Leventidis
Northeastern University
leventidis.a@northeastern.edu

Jiahui Zhang
Northeastern University
zhang.jiahu@northeastern.edu

Cody Dunne
Northeastern University
c.dunne@northeastern.edu

Wolfgang Gatterbauer
Northeastern University
w.gatterbauer@northeastern.edu

H.V. Jagadish
University of Michigan
jag@umich.edu

Mirek Riedewald
Northeastern University
m.riedewald@northeastern.edu

## ABSTRACT

Understanding the meaning of existing SQL queries is critical for code maintenance and reuse. Yet SQL can be hard to read, even for expert users or the original creator of a query. We conjecture that it is possible to capture the logical intent of queries in *automatically-generated visual diagrams* that can help users understand the meaning of queries faster and more accurately than SQL text alone.

We present initial steps in that direction with visual diagrams that are based on the first-order logic foundation of SQL and can capture the meaning of deeply nested queries. Our diagrams build upon a rich history of diagrammatic reasoning systems in logic and were designed using a large body of human-computer interaction best practices: they are *minimal* in that no visual element is superfluous; they are *unambiguous* in that no two queries with different semantics map to the same visualization; and they *extend* previously existing visual representations of relational schemata and conjunctive queries in a natural way. An experimental evaluation involving 42 users on Amazon Mechanical Turk shows that with only a 2–3 minute static tutorial, participants could interpret queries meaningfully faster with our diagrams than when reading SQL alone. Moreover, we have evidence that our visual diagrams result in participants making fewer errors than with SQL. We believe that more regular exposure to diagrammatic representations of SQL can give rise to a *pattern-based* and thus more intuitive use and re-use of SQL.

A free copy of this paper; its appendices; the evaluation stimuli, raw data, and analyses; and source code are available at https://osf.io/mycr2

## 1 INTRODUCTION

SQL is a powerful query language that has remained popular in an age of rapidly evolving technologies and programming languages. Unfortunately SQL queries are often verbose and involve complex logic constructs. This makes them hard to read to a degree where even SQL experts require considerable time *to understand* a non-trivial query.

While the difficulty of *composing* SQL queries has received much attention, it is often just as important to *read and understand* them correctly. For example, SQL queries may require maintenance as database schema or data properties evolve, or when the analysis goals change. Even the development of new queries can be facilitated by understanding and reusing existing ones. A paradigm, successfully employed in projects such as the Sloan Digital Sky Survey [74], is to begin with a query that is similar to the desired one and then modify it as needed. In fact, several systems have been proposed that let users browse and re-use SQL queries in a large repository, including CQMS [47, 48], SQL QuerIE [5, 18], DBease [53], and SQLshare [42]. The key premise of these systems is that *starting from an existing template* should make it easier to specify an SQL query than starting from scratch. However, in order for users to successfully build upon an existing SQL query, they need to understand it first.

**Our goal.** Compared to query composition, SQL query interpretation is relatively unexplored. Our goal is to provide an approach that simplifies the process of SQL query interpretation. For this purpose, we propose automatically generated *diagrammatic representations* of SQL queries that capture their logical intent. Our approach is orthogonal to SQL composition and hence can be used to complement any existing SQL development tool, whether visual or not.

**Target audience.** We target two types of users: Foremost, we like to help users who browse through a repository of existing SQL queries (e.g., their own past queries, or a log of past issued queries over a shared scientific data repository) and try to quickly understand the meaning of such queries. The purpose can be to either find a past query again, or to run queries created by others, or to study existing queries and modify them later. Our paper shows that *diagrams can speed up the process of interpreting existing SQL queries*. The second target is more speculative. We hypothesize (but do not claim to have yet evidence) that providing a formalism for SQL users to help reason in terms of *SQL patterns* can be a helpful process, both while learning SQL and also later when remembering a particular SQL pattern when composing a new query. We provide more detailed illustrating examples for the possibilities in Appendix G.

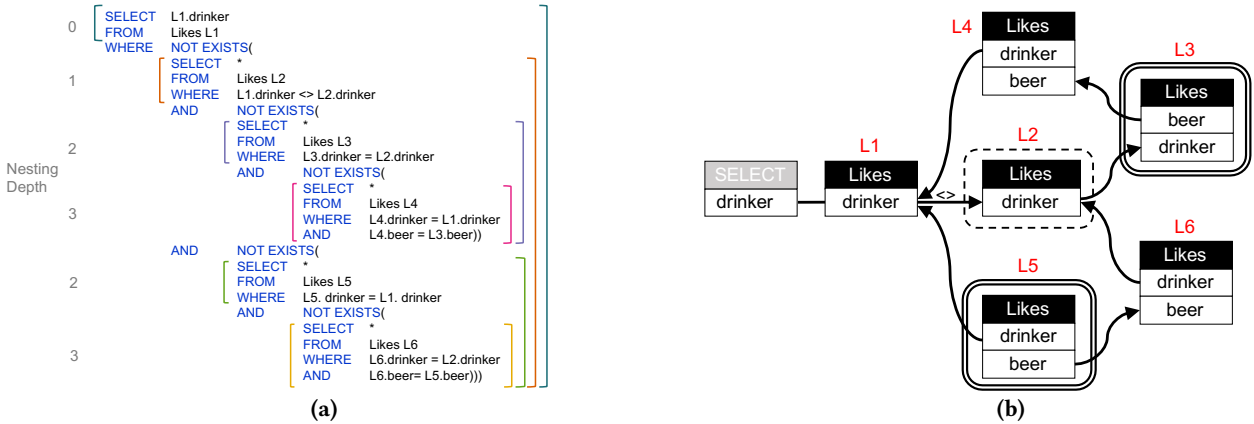

Figure 1: (a): The *unique-set-query* over the bar-drinker-beer schema, whose purpose is to *find drinkers that like a unique set of beers*. The nesting depth of each subquery is denoted in gray on the left; the scope of each subquery is shown by the brackets on the right and their respective "roots" by the brackets on the left. (b): Our visual diagram for the same query. The red table aliases next to the tables are not part of the diagram and are only placed to illustrate the correspondence to the SQL query. Notice that the visual pattern on the right is the same for different SQL queries that follow *the same logical pattern*, such as *find beers with a unique set of drinkers* or *find movies with a unique cast of actors* or *find customers with a unique set of purchased items*. Thus our diagrams allow users to inspect and recognize the underlying logical pattern.

## 1.1 Query Interpretation: An Example

SQL queries can be notoriously complex, even when they have a compact description in natural language. By visualizing the logic of an SQL query, we hope to make its logic and intent easier to understand. The following detailed example illustrates this idea.

**The unique-set query.** Consider the well-known beer drinkers schema by Ullman [78]: Likes(person,beer), Frequents(person,bar), Serves(bar,beer). Suppose we wish to find *drinkers who like a unique set of beers*, i.e., no other drinker likes the exact same set of beers. The query requires only the Likes table; its SQL text is shown in Fig. 1a. Please take a moment to look at the SQL statement and verify that it correctly expresses the desired query. If this takes you several minutes, it may not be because of your lack of SQL expertise: the logic of SQL is intricate. After some effort, the query can be read as: *return any drinker, s.t. there does not exist any other drinker, s.t. there does not exist any beer liked by that other drinker that is not also liked by the returned drinker and there does not exist any beer liked by the returned drinker that is not also liked by the same other drinker*.

**Set theory.** From a set logic perspective, this query applies the following logical pattern: Let $x$ be a drinker and $S(x)$ be the set of beers $x$ likes. Our intent is find those $x$, s.t. there does not exist another drinker $y \neq x$ for which $S(y) \subseteq S(x)$ and $S(y) \supseteq S(x)$. In other words, find *drinkers for which no other drinker has simultaneously a subset and superset of their beer tastes*. Hence, the query "merges" two logical patterns: (1) no other drinker likes a subset of the liked beers, and (2) no other drinker likes a superset of the liked beers. *The other drinker* in both logical patterns must be *the same person*.

**First-order logic.** SQL and relational calculus are based on first-order logic (FOL). FOL expresses the first pattern (no other drinker likes a subset of the liked beers) as *there does not exist any other drinker $y \neq x$, s.t. all beers liked by $y$, are also liked by returned drinker $x$*. Similarly, the second pattern (no other drinker likes a superset of the liked beers) is expressed as *there does not exist any other drinker $y \neq x$, s.t. all beers liked by a returned drinker $x$, are also liked by $y$*. Notice that both conditions must be fulfilled *simultaneously* by any other drinker $y$, thus our *composite* pattern is a *conjunction* of the two aforementioned patterns, sharing the same other drinker $y$. Also notice that SQL does not support universal quantification directly. Thus the statement *all beers liked by $x$ are also liked by $y$* needs to be transformed into the more convoluted *no beer liked by $x$ is not also liked by $y$*.

**Our visual diagrams.** Our method is a *diagrammatic representation system* based on FOL that automatically translates such logical patterns from SQL into visual patterns in a way that makes it easier for a user to inspect and recognize them.

Figure 1b shows the visual pattern for the example. A dashed bounding box represents a logical *Not Exists* ($\nexists$) and a double-lined bounding box represents a *For All* ($\forall$) quantifier, which are applied to the attributes of the enclosed tables. To read the diagram, we start from its SELECT box and follow the arrows to the next table attribute: The first pattern consists of the set of bounding boxes L1→L2→L3→L4 and reads as follows: *Return any drinker (L1), s.t. there does not exist a different drinker (L2), s.t. for all beers liked by the different drinker*

*(L3), they are also liked by the returned drinker (L4)*. The second pattern consists of bounding boxes L1→L2→L5→L6, thus sharing the first two boxes.[1] The additional conditions (forming a conjunction with the former) are: *... and s.t. for all beers liked by the returned drinker (L5), they are also liked by the different drinker (L6)*. Reading through the diagram *feels* similar to reading a FOL expression where appropriate symbols identify the predicates and the quantifiers applied to them. This proximity to FOL is not a coincidence, but rather a key feature that preserves and exposes the logic behind SQL queries, yet facilitates their interpretation. Also notice that, in contrast to SQL, we can avoid a double negation and instead use a more intuitive universally quantified statement.

Reading the query visualization may not seem simple at first. Yet notice: (1) any representation system, including classic ER or UML diagrams, may appear cryptic to a novice who sees this representation for the first time; and (2) the logic of the unique-set query is indeed non-trivial. However, because this logic is represented by only a handful of boxes, instead of multiple dense lines of SQL text, it can actually be *easier for readers to recognize*, once the visual conventions become familiar (e.g., after a short tutorial). We will present experimental results verifying this claim later in Section 6.

**Common visual patterns.** The logical pattern behind a particular query is not unique to the query, and the visual diagram remains the same for queries with identical logical patterns. For example, if we want to find all bars that have a unique set of visitors, the diagram would remain the same except for replacing table and attribute names appropriately. This is true even across schemas, e.g., for a query finding all movies with a unique cast in a movie database we obtain the same visual pattern, allowing for the recognition of similarities that are difficult to distill from pure SQL. Thus, our diagrams *expose the underlying logical patterns* to the user in a way that facilitates query interpretation and recall.

## 1.2 Challenges, Contributions, and Outline

The **challenges** we faced when developing our diagrammatic visualization were to design a representation that (*i*) can be intuitively learned and quickly understood, (*ii*) can express a large fragment of SQL, (*iii*) is not entirely detached from SQL but rather captures the essence of SQL logic, (*iv*) is minimal in that no visual element is superfluous, (*v*) is designed based on human-computer interaction best practices, (*vi*) is unambiguous in that no two queries with different semantics map to the same visualization, and (*vii*) extends previously existing visual representations of relational schemata and conjunctive queries in a seamless way.

Our **experimental study** with 42 participants shows that existing SQL users can determine the meaning of queries meaningfully faster (-20%, $p < 0.001$) and more accurately (-21%, $p = 0.15$) *using our diagrams alone* instead of standard SQL. There is also some evidence that participants make meaningfully fewer errors (-17%, p=0.16) when looking at both *our diagrams together with SQL* instead of SQL alone. These participants were recruited on Amazon Mechanical Turk (AMT) and spent only 2-3 min on a short tutorial with 6 examples of SQL annotated with their respective diagrams. Thus *while the participants had significant prior experience with SQL, they were exposed to our visualizations for only a few minutes and were still faster and typically made fewer errors interpreting the queries*. We can only imagine the improvements if users received more regular exposure to those diagrams and thereby could start to internalize the underlying *logical patterns of SQL queries*. We thus believe that our approach shows a direction that is worthwhile for our community to explore in order to make relational databases more usable [13, 46].

Our main **contributions**, in presentation order, are:
(1) We survey closely related approaches and explain why visual query builders cannot provide the functionality needed for effective query visualization (Section 2);
(2) We identify abstract visual design requirements for assisting humans in query understanding (Section 3);
(3) We present our novel diagrammatic representation of SQL, discuss its origin in first-order logic and diagrammatic reasoning systems, and justify our design choices using a theory of minimal and effective SQL visualizations (Section 4).
(4) We prove that our diagrams are *unambiguous*, i.e., it is not possible for two different logic representations to lead to the same visual diagram (Section 5);
(5) We present an empirical validation of our approach with a randomized controlled study involving 42 users, which provides evidence that existing SQL users are meaningfully faster at correctly understanding queries using our diagrams than using SQL text, despite having experienced only minimal prior training on our visualizations (Section 6).

**Earlier work and additional material.** An earlier vision paper [36] and an interactive system demonstration [25] referred to our approach as *QueryViz*, which we have since renamed to *QueryVis*. Those two short papers described the vision and a prototype implementation, yet lacked a detailed justification of the design, a proof of the diagrams being unambiguous, and an empirical user study. A full version of this paper including all appendices; supplemental materials for the user study including the stimuli, raw data, and analyses; and source code are available at osf.io/mycr2 and on the project web page queryvis.com

---

[1]The reading order follows a depth-first traversal from the SELECT box with restarts on source nodes: After the path L1→L2→L3→L4, the reading starts from L5 (that has no incoming edges) and continues L5→L6. Later Section 4.6 has the details.

## 2 RELATED WORK

For decades, SQL has been the main standard for specifying queries over relational databases and this is unlikely to change anytime soon. Thus we do not propose new ways for users to write queries, but instead explore how to help them *understand existing SQL queries*.

**Visual query languages.** Visual methods for specifying queries have been studied extensively (a 1997 survey by Catarci et al. [14] cites over 150 references) and many commercial database products offer some visual interface for users to write SQL. We focus on the problem of describing and *interpreting a query that has already been written*, which is very different from the problem of helping a user to compose a query. The central difference is that understanding a query requires a focus on the high-level structure, abstracting away low-level details and subtleties [36]. In contrast, to precisely specify a query, possible options and specific details affecting query semantics must be presented. In programming languages, this distinction is clearly made between *visual programming* for developing a program and *program visualization* for analyzing an existing program [60].

**Interactive query builders** employ visual diagrams that users can manipulate (most often in order to select tables and attributes), while using *a separate query configurator* (similar to QBE's condition boxes [86]) to specify selection predicates, attributes, and sometimes nesting between queries. dbForge [26] is the most advanced and commercially supported tool we found for interactive query building. Yet it does not show any visual indication for non-equi joins between tables and the actual filtering values and aggregation functions can only be added in a separate query configurator. Moreover, it has limited support for nested queries: the inner and outer queries are built separately, and the diagram for the inner query is *presented separately and disjointly* from the diagram for the outer query. Thus *no visual depiction of correlated subqueries is possible*. Other graphical SQL editors like SQL Server Management Studio (SSMS) [75], Active Query Builder [3], QueryScope from SQLdep [64], MS Access [56], and PostgreSQL's pgAdmin3 [63] lack in even more aspects of visual query representations: most do not allow nested queries, none has a single visual element for the logical quantifiers NOT EXISTS or FOR ALL, and all require specifying details of the query in SQL or across several tabbed views *separate from a visual diagram*. DataPlay [1, 2] allows a user to specify their query by interactively modifying a *query tree with quantifiers* and observing changes in the matching/non-matching data. The declared goal is to overcome SQL's lack of *syntactic locality* (i.e. otherwise similar SQL queries with different quantification can have very different structure). QueryVis is designed with the same issue in mind (see [36] and Fig. 2), yet also leverages familiarity with existing visual metaphors for conjunctive queries. In short, current graphical SQL editors *do not provide a single encompassing visualization of a query*. Thus they could not (even in theory) transform a complicated SQL query (such as the one from Section 1.1) into a single visual representation, which is the focus of our work.

**Query visualizations** attempt to create a visual representation of existing queries. This explicit reverse functionality for SQL has not drawn as much attention as visual query builders. The two projects that come closest in spirit are GraphSQL [15] and Visual SQL [45]. Both are visual query languages that also support query visualization. GraphSQL uses visual metaphors that are different from typical relational schema notations and visualizations, even simple conjunctive queries can look unfamiliar. Visual SQL is closer in design. With its focus on query specification, it maintains the one-to-one correspondence to SQL, and syntactic variants of the same query lead to different representations (see Fig 4 in [36]). The Query Graph Model (QGM) developed for Starburst [41] helps users understand query plans, not query intent. StreamTrace [8] focuses on visualizing temporal queries over streams with workflow diagrams and a timeline. We focus on displaying the underlying logic behind general SQL queries, independent of data.

**Information visualization** with its goal to help users *understand and analyze data* [19] has recently drawn a lot of attention in the database community. Similar to our objective, visualization researchers help users understand complex relationships, but in data instead of in query logic.

**Syntax highlighting.** Query editors and clients for major DBMSs have long used *syntax highlighting and aligning* of query blocks and clauses. This is helpful, yet not sufficient to help users understand a query's intention. In our experiments, SQL queries are auto-indented, and the keywords are capitalized and highlighted in color (see Fig. 3). Still users generally responded better to our approach than to the visually improved SQL text.

**Natural language translations.** Translating between SQL and NL is an interesting and heavily researched topic, and various ideas are proposed to *explain queries in natural language* [37, 44, 50, 73, 83]. Work in this area convincingly argues that automatically creating effective free-flowing text from queries is difficult and that the overall task is quite different from previous work on creating natural language interfaces to DBMSs. The key limitation of the current state of the art [37, 83] is that it (*i*) produces long sentences, (*ii*) is currently limited to simple SQL queries, and that (*iii*) textual descriptions do not readily reveal common logical patterns behind queries. In particular, we are *not aware of any SQL to NL tool available today* that could translate our example from Figure 1 into an intuitive NL representation.

Several papers [2, 17, 61] suggest illustrating the semantics of operators in a data flow program or the semantics of queries by generating **example input and output data**. The result is basically a list of tuples for each relational operator, which, like the natural-language explanation, does not readily reveal the logical pattern behind a query.

## 3 TASK ANALYSIS & ABSTRACTION

In order to understand the logical pattern behind a query, a user must first perform several low-level tasks. Appropriately determining what these tasks are [27] and abstracting them into domain-independent terms [12] is crucial for designing a visualization that will enable users to achieve their goals. Here we describe our task analysis and abstraction which we will later use to motivate our visualization design.

### 3.1 User Task Analysis

Users perform tasks over a visualization in order to accomplish their higher-level goal of interpreting an SQL query correctly. To determine the necessary low-level tasks we performed a scenario-based task analysis [27] based on our considerable collective experience using and teaching SQL.

In order to correctly interpret a *conjunctive query with inequalities* (e.g., the one shown in Fig. 3a) the user tasks are:
  (1) **Tables & attributes**: Identify the tables specified in the SQL query and their relevant attributes (i.e., attributes involved in a predicate or selected).
  (2) **Selection predicates**: Identify the attributes of a table that are assigned a selection predicate and the details of that predicate (e.g., T.attr1 = 4).
  (3) **Join predicates**: Identify the pairs of attributes with a join predicate as well as the join operator (e.g., T.attr1 = S.attr1, T.attr2 $\geq$ S.attr2).

We also wish to support *nested conjunctive queries with inequalities* (we discuss this SQL fragment later in Section 4.4). In order to correctly interpret such nested queries (e.g., the query in Fig. 3b), a user must additionally:
  (4) **Quantifiers**: Identify the logical quantifier ($\exists$, $\nexists$, or $\forall$) applied to a set of tables and their attributes.
  (5) **Nesting order**: Identify subqueries, their associated tables, and their the nesting order.

Addressing these additional tasks for nested subqueries adds substantial design challenges.

### 3.2 User Task Abstraction

Building on our task analysis we conducted a task abstraction [12] in order to discuss these database domain tasks using abstract visualization terms. There are three abstract properties that the visualization should clearly portray: relations, group membership, and hierarchy.

**(I) Relations.** This property includes user tasks (2) *Selection predicates* and (3) *Join predicates*. The *Selection predicates* task deals with identifying the attribute and the selection predicate. We are thus interested in identifying a relation applied to a single attribute. Similarly, the *Join predicates* task deals with identifying pairs of attributes that join tables and the associated operator for which there is a join condition.

**(II) Grouping.** This property includes user tasks (1) *Tables & attributes* and (4) *Quantifiers*. The *Tables & attributes* task deals with identifying tables and their relevant attributes that are part of the table. We are thus interested in grouping relevant attributes of a given table as well as distinguishing between tables. The *Quantifiers* task deals with identifying the set of tables to which a logical quantifier is applied. We are thus interested in collecting and grouping all the tables for which the same quantifier is applied.

**(III) Hierarchy.** The (5) *Nesting order* task deals with identifying the nesting order of subqueries and their associated tables. We are thus interested in identifying the hierarchical structure of the subqueries present in the SQL query. Note that the portrayal of hierarchy is not necessary if we only deal with conjunctive queries as every table will be at the same nesting depth (e.g., the query in Fig. 3a).

## 4 VISUALIZATION DESIGN

With our task analysis and abstraction established, we now turn to the design of our visualization.

### 4.1 Minimal & Effective Visualization

Our goal in this section is to start from two basic primitives of visualization design — *marks* and *channels* [59] — and develop from the ground up, a theory of *minimal* and *effective* SQL visualizations. We intend to demonstrate, using this theory, a more formal way for designing and evaluating visualizations for database systems.

All visualizations are composed of *marks* and *channels* [59]. *Marks* are geometric primitive objects such as points, lines, areas, and volumes. A visual *channel* is a way to control the appearance of a mark, which can be done by changing the position, shape, size, color, texture, orientation, motion path, and many other visual attributes of a given mark. With those basic building blocks for visualizations, we now would like to find visualizations that are minimal and effective:

DEFINITION 4.1. *A visualization designed for a set of user tasks is (1)* minimal *if every mark and channel used is necessary to accomplish at least one task and (2)* effective *if every mark and channel used is well-justified for the associated tasks based on human perception studies and visualization best practices.*

Note that this definition builds upon previous work in the visualization community. Our definition of effectiveness

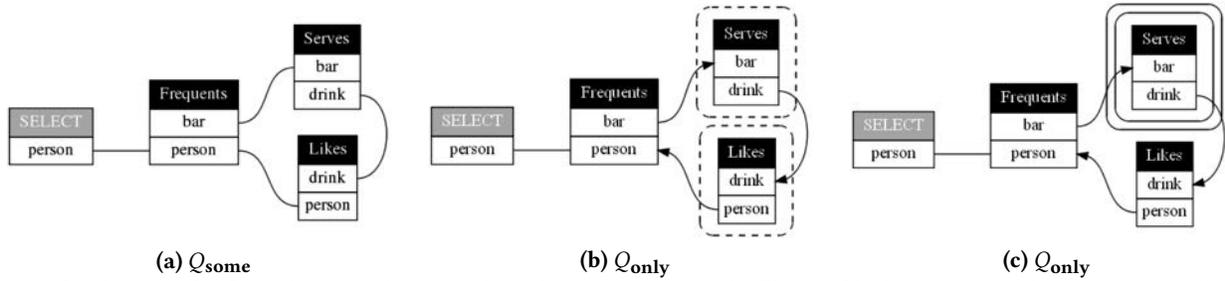

(a) $Q_{\text{some}}$    (b) $Q_{\text{only}}$    (c) $Q_{\text{only}}$

Figure 2: Section 4.8: (a) Our diagram of the conjunctive query shown in Fig. 3a. Notice how the diagram closely follows a familiar relational schema notation. (b) Our diagram of the nested query shown in Fig. 3b. Notice that we have added a dashed bounding box for $\nexists$ and the reading order can be found by following the arrows. (c) Fig. 2b can be further simplified through the use of the $\forall$ quantifier (double-lined bounding box), a logical and intuitive operator that does not exist in SQL.

```
SELECT   F.person
FROM     Frequents F, Likes L, Serves S
WHERE    F.person = L.person
AND      F.bar = S.bar
AND      L.drink = S.drink
```

(a) $Q_{\text{some}}$

```
SELECT   F.person
FROM     Frequents F
WHERE    not exists
         (SELECT   *
          FROM     Serves S
          WHERE    S.bar = F.bar
          AND      not exists
                   (SELECT   L.drink
                    FROM     Likes L
                    WHERE    L.person = F.person
                    AND      S.drink = L.drink))
```

(b) $Q_{\text{only}}$

Figure 3: (a) $Q_{\text{some}}$: Find persons who frequent some bar that serves SOME drink they like. (b) $Q_{\text{only}}$: Find persons who frequent some bar that serves ONLY drinks they like ≡ ... some bar that serves NO drink that is NO liked by them.

agrees with Mackinlay [54]. Likewise, our definition of *minimality* combines elements of Mackinlay's effectiveness and expressiveness criteria [54] and Tufte's guidance to maximize the data-to-ink ratio [77].

Like Brehmer & Munzner we define *target user tasks* as "domain- and interface-agnostic operations performed by users" [12, 58]. These tasks can be defined with varied degrees of granularity and abstraction [12, 58]. Finally, by *well-justified* we mean that sufficient evidence exists for the effectiveness and expressiveness [54] of the proposed encodings.

In the following sections we demonstrate how we designed our diagrams in the context of this theory of *minimal* and *effective* SQL visualizations. When possible we illustrate our design with figures. We refer the interested reader to our web page http://queryvis.com with an online demo of QueryVis.

## 4.2 Visual diagrams

We now formally define our visual diagrammatic representation, which we refer to as **QueryVis**, and the fragment of SQL it supports. We also describe the automatic transformation from SQL to QueryVis, formalize how to read the diagrams, and discuss some of their desirable properties.

Our diagrams are inspired by a large body of work on **diagrammatic reasoning systems** [22, 43, 72]. Diagrammatic notations themselves are inspired by the influential *existential graph notation by Charles Sanders Peirce* [62] and exploit topological properties, such as enclosure, to represent logical expressions and set-theoretic relationships. All these representation systems share an origin in that they portray *fragments of first-order-logic* (FOL) through their visual encoding, but at varying degrees of logical expressiveness and *mainly for monadic relations*. Since SQL (without grouping and null values) essentially represents a tractable FOL fragment between tables and their attributes, QueryVis focuses on capturing the FOL representation of an SQL query and representing its underlying logic visually for efficient and intuitive interpretation. We incorporate established visual metaphors from diagrammatic reasoning *into relational schemas* and adapt them where necessary or appropriate.

## 4.3 Visualizing Conjunctive Queries

We first motivate the design of our diagrams for conjunctive queries, e.g., the one in Fig. 3a and its visualization in Fig. 2a, then generalize to nested queries.

*4.3.1 Visualization Design & Effectiveness.* Based on the user tasks and abstract properties we identified in Sections 3.1 and 3.2 for conjunctive queries, we reason over the possible mark and channel choices and select the most effective one(s) for each property. Our choices are *effective* as justified using perception studies and visualization best practices.

**(1) Relations.** To portray a relation, a line mark (straight or curved) drawn between the associated attributes (several shown in Fig. 2a) can simply and clearly encode their connection [59]. In the *Join predicates* task we must distinguish between different comparison operators, which can be done efficiently by placing a label on top of a line to display the applied operator. E.g., the <> operator is shown in Fig. 1 (b) between the L1 and L2 tables. Instead of labels, one could use different line styles (e.g., dashed, double, thin). However, since there are six different operators, this would impose significant learning overhead for a new user and would be less intuitive than labels. To further minimize our design, since the most common type of join is an equijoin, we omit the = label for lines representing equijoins, i.e., unlabeled lines denote an equijoin. Moreover, since the order of elements matters for some operators such as $\{<, \leq, \geq, >\}$, we add an arrowhead mark when necessary to indicate the correct reading order (not illustrated).

For the *Selection predicates* task a line is not an effective encoding as the relation is within one element and needs no portrayal of a connection with another. A constant qualification is better portrayed *in place*, stated explicitly in a row of the referencing table which is highlighted to indicate the presence of a qualification (not illustrated).

**(2) Grouping.** There exists a plethora of ways to visualize groupings of elements [7, 79] that vary based on the amount, type, and relationship of the elements to be grouped. We adopt an explicit encoding of groups to minimize perceptual and spatial ambiguities. For the *Tables & attributes* task we want to distinguish between different tables and identify their relevant attributes. While tables may share some attributes, we deliberately consider each table as a disjoint set as we want to make the separation between tables clear. To portray this grouping we use an area/bounding box mark which allows us to use the Gestalt principle of enclosure to denote disjoint set membership [49, 59]. Since the attributes are part of a table (i.e., grouped on a per-table basis) the *table composite mark* is a fitting visual abstraction that encapsulates the membership of attributes under a table and provides a nice visual separation between different table objects. The table composite mark is made up by a set of stacked rectangular box marks, as shown in Fig. 2a. The first box/row in the table represents the table name and is filled with a black background and white text (except the SELECT table which uses a lighter background to distinguish it). The remainder of the rows display the relevant attribute names for which there is an associated selection or join predicate.

Notice that the marks and channels we have chosen, while based on our task analysis and abstraction, are in accordance with previous conjunctive query visualizations and relational schemas, i.e., they are "backwards compatible."

*4.3.2 Visualization Minimality.* Our diagrams for conjunctive queries are minimal visualizations because the removal of any mark or channel would lead to incomplete visualizations where at least one of the user tasks cannot be achieved unambiguously. Moreover, we aim to maximize the data-to-ink ratio, i.e., the proportion of a graphic's "ink" devoted to the non-redundant display of data information [77]. We now explore the issue of visualization minimality for our diagrams for conjunctive queries, noting that only 3 marks are used: table composite marks, lines with arrows/labels, and constant-qualification labels.

**(1) Table.** The table composite mark is fundamental to a diagram as it identifies the tables involved in the query and their relevant attributes. Removing it would make it impossible to interpret joins. Removing the black background of the first table row, which displays the table name, would eliminate the ability to visually distinguish the table name from its attributes. Removing the gray background in the SELECT table would reduce user ability to detect the root of the query. No attribute row may be removed either, because each must either be part of a join or a selection predicate: their removal would make the *Selection predicates* and *Join predicates* tasks impossible. Alternative visual representations could be chosen, but would require the same number of encodings.

**(2) Line.** An undirected line mark is essential for portraying joins between two attributes involving operators $\{=, \neq\}$ and its removal would make such *Join predicates* tasks infeasible. Other representation alternatives could be textual, such as referencing the joining attribute next to the other attribute on a table. However, this would lose the advantages of a visual representation where a connection can be "seen" without being carefully "read." For joins with operators $\{<, \leq, \geq, >\}$, adding an arrow to a directed line mark ensures the identification of operand order and hence is essential. As an alternative one might consider positional encoding, e.g., enforcing a left-to-right and top-to-bottom reading order. However, this is insufficient for more complicated nested queries (e.g., Fig. 1) which will be discussed in the next section. Conversely, adding arrows to each line would erroneously imply directionality.

Adding a label to the line mark for joins other than the equi-join $\{=\}$ ensures identification of the operator. Its removal would make such *Join predicates* tasks infeasible. Alternative solutions such as use of different line styles exist, but would increase learning difficulty.

**(3) Constant-qualification labels.** Without these labels the *Selection predicates* task would be infeasible. Alternative visual encodings would add clutter and learning difficulty. Removing the highlight color would limit user ability to distinguish a qualification from an attribute.

## 4.4 Nested SQL queries

Our main focus is to add nesting to conjunctive queries. We say an SQL query is *nested* if it contains at least one subquery. This is a major step as nested queries (in particular, correlated nested queries) can be very hard to interpret and are not even

```
Q::=    SELECT   C [, C, ..., C] | *        select clause
   |    FROM     S [, S, ..., S]            from clause
   |    [WHERE   P]                         where clause
C::=    [T.]A                                column or attribute
S::=    T [AS T]                             table (table alias)
P::=    P [AND P ... AND P]                 conjunction of predicates
   |    C O C                                join predicate
   |    C O V                                selection predicate
   |    [NOT] EXISTS (Q)                    existential subquery
   |    C [NOT] IN (Q)                      membership subquer
   |    C O {ALL | ANY (Q)}                 quantified subquery
O::=    < | ≤ | = | <> | ≥ | >              comparison operator
T::=                                         table identifier
A::=                                         attribute identifier
V::=                                         string or number
```

Figure 4: Grammar of supported SQL fragment. Statements enclosed in [ ] are optional; statements separated by | indicate a choice between alternatives.

supported by most visual query builders. We focus on the most expressive subqueries, i.e. those inside the WHERE clause and operators EXISTS, NOT EXISTS, IN, NOT IN, ANY or ALL. Figure 4 shows our currently supported SQL fragment. Those queries have the same expressiveness as relational calculus and its set-based interpretation. However, we currently do not support disjunctions and thus refer to this SQL fragment as *nested conjunctive queries with inequalities*. In addition, SQL queries must fulfill two minor restrictions which we define in Section 5 and argue that they are fulfilled by any meaningful *non-degenerate* SQL query.

**Notation.** A *predicate* has the form exp1 op exp2 where at most one of the exp's is a constant (e.g., 3 or 'Alice'), and the other(s) are attribute names optionally with table aliases (e.g., table1.attr2). If a predicate has a constant it is a *selection predicate*, otherwise a *join predicate*. Operator op is an element of $\{<, \leq, =, <>, \geq, >\}$. A *query block* consists of SELECT, FROM, and WHERE clauses including therein defined table aliases and predicates. We call the query block at nesting depth 0 the *root query block*. The *scope* of a query block is the set of query blocks for which the table aliases defined within it are valid (e.g., as shown by the brackets on the right of Fig. 1a). The *root of the scope* is the query block itself. Fig. 1a indicates those by brackets on the left along with their respective *nesting depths*. If a subquery has no other nested subquery then its scope and root are the same (e.g., subqueries involving table names L4 and L6 in Fig. 1a).

## 4.5 Visualizing Nested SQL queries

In order to visualize nested conjunctive queries with inequalities, we must extend the visualizations for conjunctive queries (Section 4.3.1) to further enable the *Quantifiers* and *Nesting order* tasks (Section 3.1). To this end, we are extending the conjunctive-query visualization to allow up to depth 3 nested queries. We illustrate the design for nested queries using Fig. 3b and its associated visualizations Figs. 2b and 2c.

*4.5.1 Visualization Design & Effectiveness.* We extend the design of our diagrams by keeping the same encodings for the tasks examined in Section 4.3.1 and choose the most effective additional marks and channels for the *Quantifiers* and *Nesting order* tasks from the grouping and hierarchy abstractions, respectively.

**(1) Grouping.** For the *quantifiers* we must operate a level above the *Tables & attributes* task to group tables based on a quantifier. However, we can leverage the same principles [7, 49, 59, 79] to design an area/bounding box mark that encloses a set of tables. To be distinct from table boxes we use a rounded rectangle mark. As we only have two quantifiers to encode $\{\nexists, \forall\}$, we choose to use a dashed and double line style, respectively, so as to avoid labeling. $\nexists$ dashed lines are shown in Fig. 2b while a simplified representation with $\forall$ double lines is shown in Fig. 2c.

**(2) Hierarchy.** A hierarchy can be effectively visualized as a rooted tree or similar node-link structure [59, 76]. For the *Nesting Order* task we could portray the nesting with a logic tree like we introduce in Fig. 5. This would necessitate two arrow types: one to represent the nesting order of subqueries as in logic trees and another to represent the table joins as we do for conjunctive queries (Section 4.3.1). Recall from Section 4.3.2 that positional encoding alone (also shown by Fig. 5 for logic trees) is insufficient for portraying more complicated nested queries (e.g., Fig. 1). Another approach would be to visually nest sets of tables in bounding boxes, but this would likely lead to cluttered visualizations.

Below we will show that simply by using arrow rules we provide for the reading order (Section 4.6), we can always recover the correct nesting order of each table in an SQL query (Section 5). Hence *additional marks encoding nesting would be redundant* as long as we ensure that arrows are appropriately added to the line marks for *Join predicates* to implicitly encode nesting order. Specifically, we (a) must use directed edges (lines with arrows) for equijoins of tables that are at different nesting depths and (b) determine the direction of the arrow solely by the arrow rules and not the order of attributes around an operator. As an example of the latter constraint, assume we have a join condition A.attr1 > B.attr2 where a directed line drawn from A to B denotes operator order. However, table B is a parent of table A in the nesting and thus the arrow rules state the directed edge must be drawn B→A. Thus we must rewrite the join with the equivalent condition B.attr2 < A.attr1.

*4.5.2 Visualization Minimality.* Just as we did for conjunctive queries in Section 4.3.2, we show that the our visualization is minimal also for *nested conjunctive queries with inequalities*. The removal of any of its marks or channels makes it infeasible to unambiguously achieve all the user tasks. In particular, we discuss the two additional and modified marks: lines with arrows/labels and bounding boxes.

**(1) Bounding Box.** A rounded rectangle mark encloses a query block, i.e. all tables to which a quantifier is applied. Removing it would make the *Quantifiers* task infeasible. One alternative to using enclosure would be to attach the quantifier and a block label or color to each table, but this would require the addition of more marks/channels and would not provide the user with the *at-a-glance* grouping of enclosure.

**(2) Line arrows/labels.** Besides identifying the pair of attributes involved in a join to support the *Tables & attributes* task, the lines and their arrows/labels are also required for the *Nesting order* task. Removing the lines or their component arrows/labels would make these tasks infeasible. However, the arrow rules (Section 4.6) can be used to unambiguously determine the nesting order of the subqueries in the SQL query with this limited additional encoding as we show in Section 5.2. Alternative approaches, such as positional encoding or nested enclosure, would be insufficient in many cases, and would likely require many more marks and thus more "ink" [77] to display the same data.

### 4.6 Reading Order

QueryVis diagrams are read by starting from the SELECT table and following a depth-first traversal with restarts from unvisited source nodes (i.e. those without incoming arrows).

Assume an edge goes from S.attr1 to T.attr2, there is a quantifier $\nexists$ applied to T and the edge is labeled with comparison operator <. Then we can interpret that as follows: *Find* attr1 *from* S *s.t. there does not exist any tuple in* T *where* S.attr1 < T.attr2. Recall that unlabeled edges represent equijoins. Also notice that if the two tables are from the same query block, then they are treated as if T has the $\exists$ quantifier applied. Once we finish interpreting an edge, we need to add an AND in our interpretation because we represent a conjunction of predicates.

### 4.7 Transforming SQL into diagrams

Here we provide an overview of the diagram creation process. We first convert an SQL query into tuple relational calculus (TRC), which is a well-studied transformation to FOL [21]. As a consequence of transforming into FOL, we consider set semantics, 2-valued logic (no NULLs), and no aggregate functions. Moreover, we connect multiple predicates only by conjunctions (i.e., no disjunctions are allowed). In FOL we *no longer have to deal with the various syntactic variants of SQL operators* which do not add expressiveness. This means that operators such as IN, NOT IN, or ALL would be converted to the corresponding FOL quantifiers $\exists$, $\nexists$ or $\forall$.

**Logic Tree (LT).** Instead of using the TRC representation of a query, it becomes easier to reason over an equivalent representation that makes the *nested scopes of the quantifiers* explicit in the form of a tree, which we denote as *Logic Tree* (LT). It is a rooted tree with each node representing a query block. The *root node* represents the root query block, and the tree structure encodes the nesting hierarchy, i.e., tables and attributes of a node can be referenced in any subtree. Each node in the LT holds the following information:

(1) *Tables (T)*: The set of tables (or table aliases) defined in its root of the scope;
(2) *Predicates (P)*: The set of predicates used in the query block. Multiple predicates in the set are related by a conjunction (i.e., predicate1 ∧ predicate2 etc.);
(3) *Quantifier (Q)*: The quantifier applied to the predicates including its negation. This is either $\exists$, $\nexists$ or $\forall$.

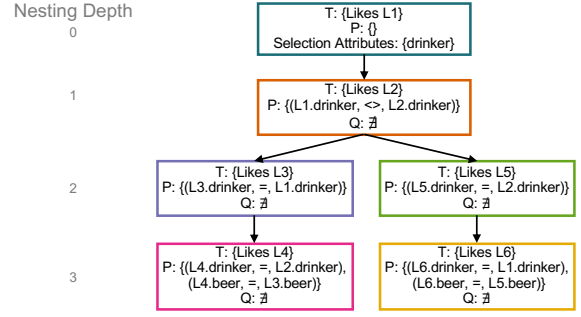

Figure 5: LT representation of the SQL query from Fig. 1a.

Moreover, for the root node of the LT, we also specify the attributes in its select-list (see Fig. 5).

**Logic Simplifications.** In SQL, queries with universal quantifiers (e.g., our example from Section 1.1) are expressed through nested NOT EXISTS subqueries, which makes them difficult to read. We simplify a LT with nested $\nexists$ quantifiers by applying a standard logical transformation to turn them into $\forall$ quantifiers. In particular, if a LT node $\psi$ has $\nexists$ as its quantifier and only has one child node $\psi'$ that also has $\nexists$ as its quantifier then we can transform $\psi$ to have a $\forall$ quantifier and $\psi'$ to have an $\exists$ quantifier.

To see why, consider following transformation where we apply De Morgan's law $\neg \exists x \in X.\big(P(x)\big) \equiv \forall x \in X.\big(\neg P(x)\big)$ with $P(x)$ being a propositional logic expression and $\neg a \lor b \equiv a \rightarrow b$ [33] on two LT nodes. Here we write T and S for a set of tables in two query blocks:

$$\neg \exists S.(p_1 \land \cdots \land p_k) \land \neg \exists T.(p_{k+1} \land \cdots \land p_{k+\ell}) \quad (1)$$
$$\forall S.\neg\big((p_1 \land \cdots \land p_k) \land \neg \exists T.(p_{k+1} \land \cdots \land p_{k+\ell})\big) \quad (2)$$
$$\forall S.\big((p_1 \land \cdots \land p_k) \rightarrow \exists T.(p_{k+1} \land \cdots \land p_{k+\ell})\big) \quad (3)$$

As example, applying this transformation to the LT shown in Fig. 5, led to our diagram from Fig. 1b.

**Our Diagram (QueryVis).** We then transform an LT into our diagram. Starting from the root node we traverse the LT tree in a breadth-first manner and we process each node as follows: (1) Create a table with its attributes for each table defined in the node. (2) Create an appropriate bounding box over the created tables based on the quantifier applied to the node. (3) Update the tables to express the selection predicates by directly writing them into a new row in the table. (4) Create arrows (directed edges) between attributes of tables for which there is a join predicate.

The interesting part is *determining the arrow direction* in step (4): If the join predicate is between two tables $T_1$ and $T_2$ from two different LT nodes $\psi_1$ and $\psi_2$, respectively, then we resolve the arrow direction as follows:

(1) If node $\psi_1$ is a parent of node $\psi_2$ then the arrow points from an attribute in $T_1$ to an attribute in $T_2$;
(2) Else the arrow points from an attribute in $T_2$ to an attribute in $T_1$

Finally, label the arrow based on the join operator op unless op is "=". As we show in Section 5.2 and Section 3, this seemingly arbitrary choice of arrow directions allows us to create diagrams that (*i*) are *unambiguous w.r.t. their logic meaning*, (*ii*) are *minimally verbose*, and (*iii*) have a *natural reading order* implied by the arrows (cp. to Fig. 1b).

## 4.8 Diagram Properties

Our diagrams are designed to enhance understanding of an existing SQL query, not to create a query from scratch or to debug it. Hence they provide the user with simple visual abstractions that ensure the following properties:

**(1) Existing metaphors as starting point.** Most database users have seen relational schema diagrams before. A simple conjunctive query should not be visualized much differently from a database schema representation. We started from UML and its familiar elements for data modeling and then added visual elements only as needed. *Conjunctive queries* such as Fig. 3a have the lowest visual complexity and are represented as shown in Fig. 2a. In contrast to the SQL representation of the query, in our diagrams *no aliases are needed*. Notice how the lines between the different attributes visualize the join operation and its conditions.

**(2) Minimal visual complexity.** QueryVis uses a minimal number of visual elements (text, line, arrow, table and bounding box) as we describe in greater detail in Section 3. Since our goal is query interpretation, not query specification, we are able to abstract away details such as NULL value handling, which do not impact the intent of the query.

**(3) First-order logic representation.** QueryVis adapts visual metaphors from diagrammatic reasoning into relational schemas. Fig. 3b shows a more complex nested query; its diagram is shown in Fig. 2b. The natural language translations of this query and of the conjunctive query in Fig. 3a have approximately the same length. Similarly, Fig. 2b shows an only slightly increased visual complexity (13% more visual elements) over Fig. 2a. In contrast, the SQL text is much more complex (167% more words), which is a known deficiency of SQL that was aptly termed *poor syntactic locality* [2] and pointed out earlier [36]. We further simplified the visual representation by adding a universal quantifier (∀), a construct that does not exist in SQL. For example, the representation from Fig. 2b can be further simplified to the one in Fig. 2c by applying the transformation described in Section 4.7. This representation now has only 7% more visible elements than the conjunctive query from Fig. 2a.

**(4) Reading-order.** Another concept borrowed from diagrammatic reasoning is a *default reading order* [34]. Note from Fig. 2c how the arrows between the relations correspond to the natural language translation. Without arrows, there would be no natural order placed on the existential and universal quantifiers.

## 5 UNAMBIGUOUS DIAGRAMS

We show that QueryVis diagrams for practically relevant SQL queries are provably *unambiguous*, i.e., every diagram can be mapped back to a unique LT. This property is crucial as otherwise a diagram could allow different interpretations.

By construction, each LT node's content (table aliases, predicates, quantifier) can be directly inferred from the corresponding tables and their connections and bounding boxes in the diagram. Hence we only have to show that all parent-child relationships between LT nodes can also be recovered. This is not obvious, because we decided to forego an explicit encoding of the LT hierarchy in the QueryVis diagram to keep the visual structure simple and to make our diagrams easier to read. Not surprisingly, our supported SQL fragment is powerful enough to create queries that have structurally different LTs, but which map to the same diagram. On the other hand, for each meaningful SQL query we observed in practice it is possible to map its QueryVis diagram back to a unique LT. We will formalize the common properties of these practical queries, which we call *non-degenerate*.

## 5.1 Non-degenerate SQL Statements

Consider again the unique-set query in Fig. 1a and observe the following properties. First, each join condition references a relation whose alias appears in the FROM clause of the same query block. For instance, L1.drinker <> L2.drinker appears in the block introducing L2, even though it could in theory also appear at a deeper nesting depth. The latter does not make sense in practice, as the user will attempt to define a condition "where it belongs," i.e., as early as possible. Second, note how each table alias introduced at shallower depth is referenced by a join condition at a deeper level. This again makes sense, because the main purpose of introducing a complex nesting structure is to relate attributes from an outer block with attributes in nearby inner blocks.

We formalize these commonly observed properties below and then show that queries with those properties result in unambiguous QueryVis diagrams.

PROPERTY 5.1 (LOCAL ATTRIBUTES). *Each predicate in a query block references at least one* local *attribute, i.e. an attribute of a table from the same query block.*

Property 5.1 implies that the predicate is placed as high as possible in the nesting hierarchy of the query and it is not possible to pull it up to an ancestor in the LT. If this property does not hold, then the predicate violating this property actually expresses a *disjunction*. For example, consider the following query written in TRC:

$$\{F.person \mid F \in Frequents \land$$
$$\neg \exists S \in Serves.(S.bar = F.bar \land F.bar = \text{``Owl''})\}$$

This query is in the SQL fragment of Section 4.4 but violates Property 5.1: the selection predicate $F.bar=\text{``Owl''}$ could be pulled up to nesting depth 0, and after applying De Morgan's law on the expression we get a disjunction:

$$\{F.person \mid F \in Frequents \land$$
$$(F.bar \neq \text{``Owl''} \lor \neg \exists S \in Serves.(S.bar = F.bar))\}$$

PROPERTY 5.2 (CONNECTED SUBQUERIES). *Each nested query block $q_i$ either has a predicate referencing an attribute from its parent query block, or each of its directly nested query blocks references both $q_i$ and its parent.*

Property 5.2 ensures that there are meaningful logical connections between a query block and its nested subqueries.

## 5.2 Proof of unambiguity

In addition to being non-degenerate, the queries we observe in practice also do not have more than 3 levels of nesting. Hence we call a diagram *valid* if there exists a non-degenerate SQL query up to nesting depth 3 that is part of the SQL fragment discussed in Section 4.4 that maps to it. We next show that any valid diagram can be uniquely interpreted.

PROPOSITION 5.1 (UNAMBIGUITY). *For any valid QueryVis diagram there exists exactly one LT that maps to it.*

We confirm Proposition 5.1 by considering all possible valid diagrams up to nesting depth 3. Given the arrow rules in Section 4.6 one can uniquely identify the corresponding LT node for any given query block in our diagram.

## 6 EXPERIMENTAL EVALUATION

We designed a user study to test whether our diagrams help users understand SQL queries *in less time* and *with fewer errors*, on average. We thus tested two **hypotheses**:

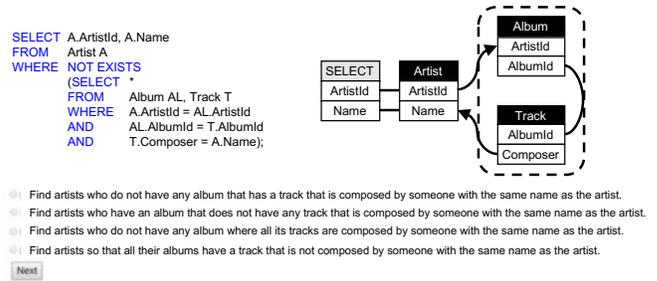

Figure 6: Example query from our study. The query is shown in the *Both* condition, in which a participant sees the query in both SQL (left) and our QV diagram (right).

**(H1)**: Study participants can understand queries *in less time* with our diagrams than by reading SQL code alone.
**(H2)**: Study participants can understand queries *with fewer errors* with our diagrams than by reading SQL code alone.

The study design and analysis plan was preregistered before we conducted the experiment and is available on OSF at osf.io/vy2x3 The complete study materials and data are also available on OSF at osf.io/mycr2

### 6.1 Study Design

To test our hypotheses, we designed an easily-scalable *within-subjects study* (i.e., all study participants were exposed to all query interfaces [69]) and made it available for 3 weeks from Jan 24, 2020–Feb 13, 2020 on Amazon Mechanical Turk (AMT). During that time, we recruited $n = 42$ legitimate participants. In the study, we presented 9 queries to participants in one of 3 conditions: (1) seeing a query as SQL alone (*SQL*), (2) seeing a query as a logical diagram that was generated from SQL (*QV*), or (3) seeing both SQL and QV at the same time (*Both*). We then tracked the time needed and errors made by each participant while trying to find the correct interpretation for each query. Note that the entire study included 12 queries yet 3 of them are related to an extension of our visualization under development (Groub-By queries) and are not presented here. The results of analyzing all 12 questions are similar to analyzing only the 9. Please see our supplemental material at osf.io/mycr2 for details.

**Multiple-choice questions.** Our study consisted of 9 multiple-choice questions (MCQs) $Q_1$–$Q_9$. Each MCQ asked the participant to choose the best interpretation for a presented query from four choices. Following best practices in MCQ creation [85], all 4 of the choices were designed to read very similar to each other so that a participant with little knowledge of SQL would be incapable of eliminating any of the 4 choices. Upon answering a question we would provide immediate feedback to the participant by highlighting the correct answer. All 9 of our questions were based on the widely used Chinook database schema [20]. Our 9 questions were split into 3 categories: conjunctive with no self-joins, conjunctive with self-joins and nested queries where each category had 3 questions (3 questions fit well in our Latin square design of 3 conditions). We chose those 3 categories based on past work [4, 11] that analyzes the most prevalent types of SQL errors. Each category consisted of 3 queries: one simple, one medium and one complex. The complexity was designated based on the number of joins and number of table aliases referenced in the query. Figure 6 shows the interface for the condition *Both* for one of the 9 questions.

**Latin square design.** For our within-subjects study we adopted a *Latin square design* [52, 57] by which each participant experienced all three conditions (SQL, QV, and Both) in a particular randomized way that reduces potential biases in our analysis due to condition ordering effects. Each participant answered *all 9 questions in the same order* but their condition for each question was different and depended on their *sequence number* in $\{S_1, S_2, \ldots, S_6\}$. The sequence number is the order in which conditions appear and is based on a repeating triplet of conditions. There are in total 6 sequences, one for each of the 6 possible permutations in a triplet (e.g., sequence $S_1$ is "SQL → QV → Both", and $S_2$ is "SQL → Both → QV"). Once a participant completes the first triplet of questions the permutation repeats. As a result, *each participant experiences each condition in 3 different questions* (e.g., a participant in sequence #1 sees questions $\{Q_1, Q_4, Q_7\}$ presented as SQL). We assigned a sequence number to each participant in a round robin fashion and ensured a balanced number of participants in each sequence.

**Recruitment of study participants.** Upon approval from our Institutional Review Board, we recruited participants from *Amazon Mechanical Turk* (AMT). Informed consent was gained from all participants before they accepted the task. We restricted participants to be from the USA, since those workers have literacy in English (which is the language our test was written in), and the USA is the country with the largest number of AMT workers [28]. We also restricted participants to those with a 95% *approval rating* on AMT, which is a typical requirement in order to reduce potential speeders (i.e., Turk workers that speed through tasks in the hopes of fulfilling the minimum requirements). All our study participants had to pass an *SQL qualification exam* to ensure that they had at least a basic proficiency with SQL. The qualification exam consisted of 6 SQL questions, took at most 10 min, and workers needed at least 4/6 correct answers.

We conducted a $n = 12$ pilot study followed by the full study with $n = 42$ AMT workers. An additional 38 *illegitimate* workers were excluded from the analysis because of obvious speeding or cheating behavior (see the supplemental material for details). The 42 *legitimate* participants included 15 women, 24 men, and 3 that didn't disclose their gender. Participants came from a variety of different professions ranging from software engineers and college students to homemakers. Each worker could only take the test once and we prevented workers from taking the test if they had participated in our pilot studies.

**Visual diagram tutorial.** Once workers passed the qualification exam and accepted our task, they were given a self-paced six-page tutorial about our logical diagrams. The tutorial introduced our basic visual notations by showing SQL examples and their diagrams. The mean (resp. median) time spent on the tutorial was approx. 3 (resp. 2) minutes. Note that our participants potentially had many years of experience with SQL[2], but *only a few minutes of experience with our logical diagrams*. The tutorial is available in our supplemental material.

**Performance-based monetary incentivisation.** Once they finished the tutorial, participants were required to answer at least 5 test questions correctly within 50 minutes in order for their work to be accepted and for them to earn a base pay of $5.20 USD. The time limit was placed in order to encourage participants to take the test in one sitting. This is critical because one of our performance metrics is the *participant's time spent per question*, and it is not uncommon for AMT workers to accept multiple tasks with long time limits at the same time and do them in parallel [16]. The base pay was determined following standard guidelines [81] based on mean pilot test duration and a USA living wage of $15/hr. To incentivize the workers to complete the test *efficiently and correctly* we used a staggered monetary incentivation scheme that added additional bonuses for having more questions correct in shorter time. Please see the online materials on OSF at `osf.io/mycr2` for details.

## 6.2 Study Analysis

We conducted a power analysis with timing data from a pilot study to estimate the required number of participants. We then used traditional null hypothesis significance testing with corrections for multi-hypothesis testing [9] as well as interval estimation of effect sizes [24, 29].

**Required sample size through power analysis.** In order to test *the effect of each condition on time*, we ran a pilot and conducted a power analysis [57, 84] to *estimate the number of participants we needed*. We did not do power analysis on error due to the discrete nature of the data. The power of a statistical test is the pre-study probability of correctly rejecting the null hypothesis, e.g., detecting a difference between conditions when one exists [39]. The rejection threshold associated with the $p$-value is denoted $\alpha$ and the power is denoted $1 - \beta$. Our power analysis assumes comparing two-sample means with a one-tailed test given parameters of $\alpha = 5\%$ and $1-\beta = 90\%$. Our two-sample means were the measured *mean times* for conditions *SQL* and *QV* based on the $n = 12$ pilot study. We conducted a one-tailed test with the alternative hypothesis that the mean time of the *QV* condition is less than the mean time of the *SQL* condition, in contrast to the null hypothesis which was the reverse statement. We chose not to use mean error results in our power analysis as the values are discretized due to the small number of questions per condition and thus unsuitable for a one-tailed test. Based on the pilot, the estimated sample size required to achieve the desired power was $n = 84$, rounded up to the nearest multiple of six to ensure an even split of participants across sequences. During the 3 weeks the study was available on AMT, we were only able to recruit 42 legitimate participants for which we report results below.

---

[2]In a post-test survey, the median SQL usage frequency was 2 times per month

**Statistical tests for rejecting the null.** We calculated 6 quantities: The mean time per question in seconds and the mean error rate in [0, 1], across all three conditions. We had 2 hypotheses for time and 2 hypotheses for error:

**Time**: (1) $\text{time}_{QV} < \text{time}_{SQL}$, and (2) $\text{time}_{Both} < \text{time}_{SQL}$.
**Error**: (1) $\text{err}_{QV} < \text{err}_{SQL}$, and (2) $\text{err}_{Both} < \text{err}_{SQL}$.

As we conducted two hypothesis tests on the same data (i.e., two hypotheses on time, and separately two on error), we *adjusted all p-values using the Benjamini and Hochberg procedure* [9] in order to minimize false discoveries caused by multiple hypothesis testing. We examined the distributions of the collected data to ensure that the assumptions of our statistical tests are not violated. We tested the normality of the distribution for each condition using a Q–Q plot and Shapiro-Wilk test [70] with $\alpha = 5\%$. We determined that the data was not normally distributed, as is required for a *parametric statistical test* [71]. Moreover, the distributions were not all transformable to normal using the same exponent via a Box-Cox transformation [68]. Thus, we used *non-parametric statistical tests* [71] which do not require our data to be normally distributed. We separately ran one-tailed Wilcoxon signed-rank [82] tests for our four hypotheses.

Notice that, as a within-subjects-study, we are comparing how the performance (time and error rate) of a participant varied between the 3 conditions. This analysis is done *for each participant individually* before averaging across participants.

**Interval estimation of effect size.** Estimation involves reporting effect sizes with interval estimates, enabling more nuanced interpretations [24, 29]. We augment our analyses with estimation because traditional dichotomous inference [10] or dichotomania [38] in null hypothesis significance testing — e.g., using only $p < 0.05$ as a cutoff for "significant" results — can lead to erroneous beliefs and poor decision making [10, 38, 39, 80]. A large $p$-value does not imply evidence in favor of the null hypothesis. Instead, a $p$-value should be interpreted as how compatible the null hypothesis is with the data and any $p < 1$ indicating some incompatibility [39, 80].

In particular, we report the sample median completion time and sample mean error and use the difference between the sample means/medians as the effect size [23, 30]. We use bias-corrected and accelerated (BCa) *95% confidence intervals (CIs)* to indicate the *range of plausible values* for the mean time and mean error [29, 31].

## 6.3 Study Results and Discussion

The results of our study on the 42 legitimate participants are summarized in Fig. 7 and the red boxes below.

> *Time.*
> (1) There is strong evidence that participants are meaningfully faster (-20%) using QV than SQL ($p < 0.001$)
> (2) Participants take a similar amount of time (-1%) using Both and SQL ($p = 0.30$)

> *Error rate.*
> (1) There is weak evidence that participants make meaningfully fewer errors (-21%) using QV than SQL ($p = 0.15$)
> (2) There is weak evidence that participants make meaningfully fewer errors (-17%) using Both than SQL ($p = 0.16$)

Note that the $p$-values above are adjusted to account for multiple hypothesis testing.

**Validation, speeders, cheaters.** In our study design and preregistration, we foresaw the problem of speeders. However, we discovered that some users cheated. Near the end of the study, many participants had all questions correct but needed less than 30 seconds to solve them. We excluded 38 users we believe cheated in our study (see supplemental material for details).

**Future potential from regular use.** Our user study provides strong evidence that our novel logical diagrams helped participants interpret queries more quickly than using SQL alone, and good evidence that combining our diagrams together with SQL helped them interpret queries with fewer errors. We would like to emphasize that the study participants had only a few minutes of exposure to our logical diagrams, but potentially many years of experience with SQL. Moreover, from a post-test survey about their frequency of SQL use, the median reported frequency was a couple times per month. We may further reduce user time and error rate by providing an interactive rather than static tutorial [51].

**Logical SQL patterns.** Since the notion of *patterns* was formulated by the architect Christopher Alexander in the 1970s [6], *patterns, archetypes, frameworks, templates* have been successfully used in various sciences, e.g., the widely-used *design patterns* in software engineering [35]. Despite having taught SQL to hundreds of student, we are not aware of any current pedagogic approach for systematically teaching SQL using *logical SQL design patterns*. We are also not aware of any ongoing efforts to apply the rich literature on logical diagrams to the task of abstracting SQL queries using *logical SQL patterns*, which may help users quickly understand the semantics and meaning of queries. We speculate that for understanding query intent, *the potential to further improve speed and reduce error rates through learning and regularly using logical SQL patterns is enormous*. We hope that further studies can verify this hypothesis. We also hope that our preliminary results can serve as a call to action for the database and visualization communities to investigate novel design and education approaches for *understanding SQL queries and their patterns*.

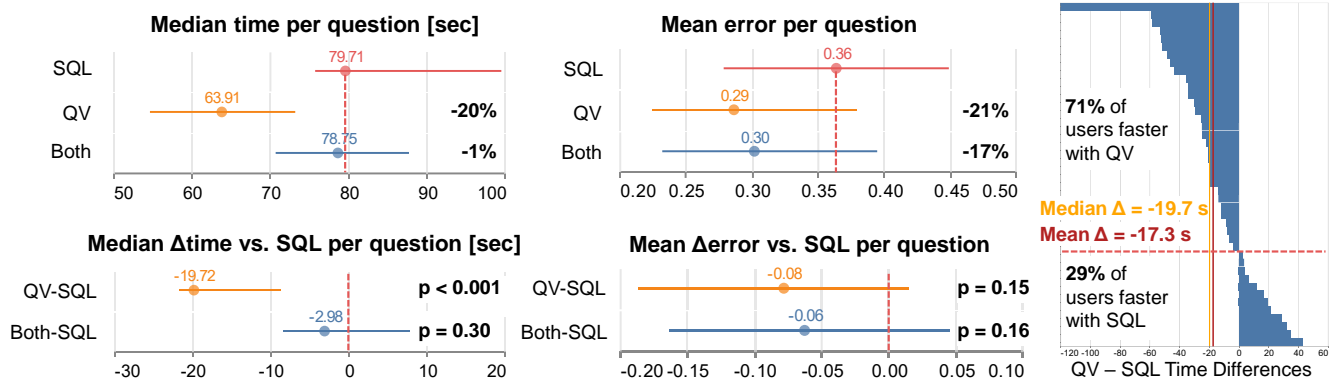

Figure 7: Median time per question and mean error rates for each of the 3 conditions with their respective 95% BCa [31] confidence intervals (CIs). Participants were on average faster using QV than SQL (-20%, $p < 0.001$), and slightly faster using Both than SQL (-1%, $p = 0.30$). Participants made less errors using QV than SQL (-21%, $p = 0.15$) and less errors with Both than SQL (-17%, $p = 0.16$). Note that the $p$-values are based on within-subjects statistical tests, thus confidence intervals of the mean could overlap substantially without indicating a high $p$-value [29]. The $p$-value for a given hypothesis tests an individual's performance in the QV or Both condition vs. their performance in the SQL condition. The $p$-values are listed along with the respective percentage difference of the condition's values vs. their counterpart SQL value. However, the confidence intervals capture the median/mean of a given condition for all participants — neglecting analysis at a within-participant level. The bottom row shows the per-participant differences of QV or Both condition vs. their performance in the SQL condition.

## 7 CONCLUSIONS

SQL queries can be challenging to read and understand, even for experts. Code maintenance, query modification, and query reuse require users to *read existing SQL queries*, both their own and those written by others. To facilitate this task, we have described an approach that can automatically transform a large fragment of SQL queries into *unambiguous*, and *visually minimal* visual diagrams. Those diagrams are based on a first order logic interpretation of SQL and thus *reflect and expose the logical patterns* innate to SQL. Our diagrams significantly improve the speed of understanding SQL queries even with virtually no prior exposure to our diagrams by participants. In future work we will investigate visual metaphors and extend the supported fragment of SQL, in particular for groupings and arithmetic predicates.

## ACKNOWLEDGMENTS

This work was supported in part by a Khoury seed grant program, the National Science Foundation (NSF) under award numbers CAREER IIS-1762268 and ACI-1640575, and by U.S. ONR under OTA N000141890001. We also like to thank Jonathan Danaparamita for singlehandedly implementing the original QueryVis prototype.

# A  CONSTRUCTING A QUERYVIS DIAGRAM

In this section we will demonstrate through a complete example (i.e., the unique drinkers example query from Fig. 1a) of how we can construct a QueryVis diagram from an SQL query in the 4 steps illustrated in Fig. 8.

## A.1  Tuple relational calculus

Given a valid SQL query (i.e., an SQL query that obeys the properties as outlined in Section 5.1) we can convert it into tuple relational calculus (TRC) [21]. This is a standard process that is well-studied and taught in advanced database classes [65]. The TRC is converted to a logic tree which essentially encodes the same information as the TRC expression but utilizes a tree structure to portray nestings whereas TRC uses parentheses/brackets.

Fig. 9a shows the TRC representation of the SQL query in Fig. 1a. Figure 10a shows the equivalent logic tree.

## A.2 Logical simplifications

Once our SQL query is in TRC form we could optionally simplify it by replacing nested $\not\exists$ quantifiers into $\forall$ quantifiers. The details of those transformations are discussed in Section 4.7.

Figure 9b shows the TRC expression after it has been simplified, and Fig. 10b shows the equivalent logic tree. Simplifying the TRC expression is optional.

## A.3 QV diagrams

Having a LT representation of the SQL query the QueryVis diagram can be finally constructed. We perform the following set of steps to obtain our QueryVis diagram:

(1) For each node in the LT we create a table in our QueryVis diagram for each table defined in the current node. The set of QueryVis diagram tables created for a given LT form a *table group*. In our case the LT from Fig. 10a has for every LT node a single table defined in it. So we create 6 tables in our QueryVis diagram and denote them as L1, L2, L3, L4, L5, L6. Consequently we also have 6 table groups.
(2) Create an appropriate bounding box over each table group based on the representing quantifier in the LT. In our case every LT node from Fig. 10a has the $\not\exists$ quantifier except the root node. We use a dashed-line bounding box to represent the $\not\exists$ quantifier. So we create a dashed-line bounding box over the table groups {L2}, {L3}, {L4}, {L5} and {L6}.
(3) Update the tables in the QueryVis diagram in order to express the selection predicates/attributes referenced in the LT. The tables are updated by directly writing the predicate/attribute into a new row in their respective table. In our case tables L1 and L2 have attribute drinker referenced (L1.drinker, L2.drinker) so we add a row with drinker in tables L1 and L2. Tables L3, L4, L5, and L6 have attributes drinker and beer referenced so we add two rows; one with drinker and one with beer for tables L4, L5, L5, and L6.
(4) Create arrows (directed edges) between attributes of tables for which there is a join predicate and label the accordingly. The arrow direction is determined by the the nesting depth relationship of the two connecting

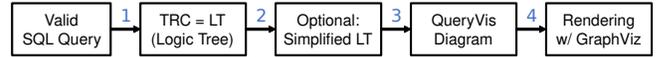

Figure 8: Flowchart of the QueryVis diagram construction pipeline. Given a valid SQL query we transform it into its tuple relational calculus (TRC) expression which is equivalent to the more intuitive logic tree representation. We can optionally simplify using our logic simplifications discussed in (Section 4.7). We then construct our QueryVis diagram from the logic tree, which is then rendered with the help of GraphViz [32].

tables. If there is a join predicate between two tables that are a depth of one apart, then the arrow points from the table with the lower depth the the table with the greater depth. If the difference in depth is greater than one, then the arrow points from the table with the greater to the one with the lower depth. If they have the same depth then the edge is undirected. The edge needs to be labeled with the comparison operator if the join was on an operator other than equality (=).

In our case we have to draw an edge for each of the following joins {(L1.drinker, <>, L2.drinker), (L3.drinker, =, L1.drinker), (L5.drinker, =, L2.drinker), (L4.drinker, =, L2.drinker), (L4.beer, =, L3.beer), (L6.drinker, =, L1.drinker), (L6.beer, =, L5.beer)}.
- For the edge between L1.drinker and L2.drinker the arrow points from L1 to L2 since L1 is at depth 0 and L2 at depth 1. We also label the edge with '<>' because the join is not an equijoin.
- For the edge between L3.drinker and L1.drinker the arrow points from L3 to L1 since L1 is at depth 0 and L3 at depth 2.
- For the edge between L5.drinker and L2.drinker the arrow points from L2 to L5 since L2 is at depth 1 and L3 at depth 2.
- For the edge between L4.drinker and L2.drinker the arrow points from L4 to L2 since L2 is at depth 1 and L4 at depth 3.
- For the edge between L4.beer and L3.beer the arrow points from L3 to L4 since L3 is at depth 2 and L4 at depth 3.
- For the edge between L6.drinker and L1.drinker the arrow points from L6 to L1 since L6 is at depth 3 and L1 at depth 0.
- For the edge between L6.beer and L5.beer the arrow points from L5 to L6 since L5 is at depth 2 and L6 at depth 3.

Another way to quickly visualize all the above arrow directions is by drawing directed edges between the logic tree nodes. We illustrate this in Fig. 11 where each table group is connected with each other based

$\{Q \mid \exists L1 \in Likes\ [L1.drinker = Q.drinker \land$
$\nexists L2 \in Likes\ [L2.drinker <> L1.drinker \land$
$\nexists L3 \in Likes\ [L3.drinker = L2.drinker \land$
$\nexists L4 \in Likes\ [L4.drinker = L1.drinker \land L4.beer = L3.beer]] \land$
$\nexists L5 \in Likes\ [L5.drinker = L1.drinker \land$
$\nexists L6 \in Likes\ [L6.drinker = L2.drinker \land L6.beer = L5.beer]]]]\}$

(a)

$\{Q \mid \exists L1 \in Likes\ [L1.drinker = Q.drinker \land$
$\nexists L2 \in Likes\ [L1.drinker <> L1.drinker \land$
$\forall L3 \in Likes\ [L3.drinker = L2.drinker \land$
$\exists L4 \in Likes\ [L4.drinker = L1.drinker \land L4.beer = L3.beer]] \land$
$\forall L5 \in Likes\ [L5.drinker = L1.drinker \land$
$\exists L6 \in Likes\ [L6.drinker = L2.drinker \land L6.beer = L5.beer]]]\}$

(b)

Figure 9: Tuple relational calculus (TRC) representation of the unique drinkers query from Fig. 1a. (a) shows the TRC representation without simplification and (b) shows the TRC expression that has been simplified by replacing the $\nexists$ quantifiers for tables L3 and L5 with $\forall$ quantifiers and the $\nexists$ quantifiers for tables L4 and L6 with $\exists$ quantifiers.

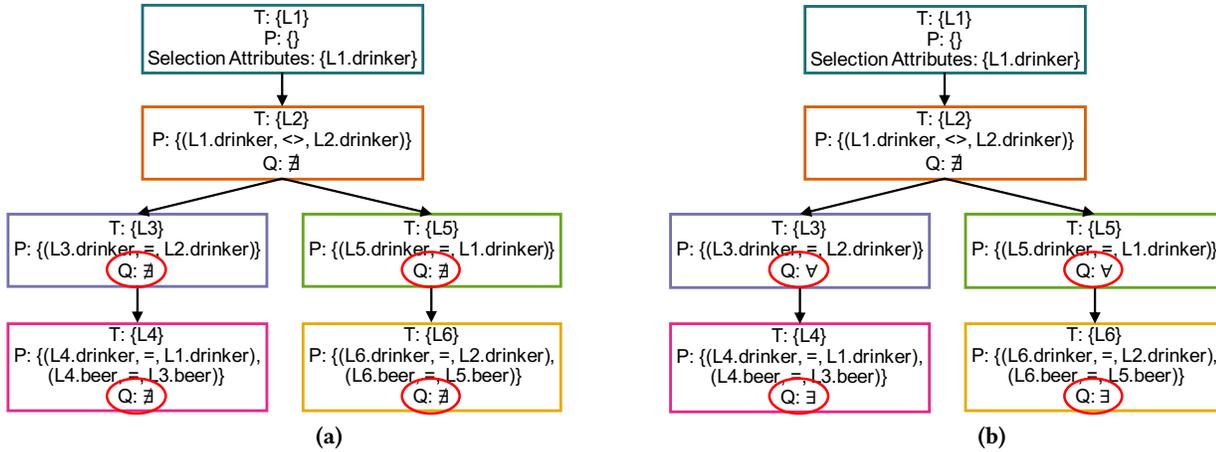

Figure 10: Logic tree representations based on the query from Fig. 1a. (a) was constructed from the TRC expression shown in Fig. 9a and (b) was constructed from the TRC expression shown in Fig. 9b.

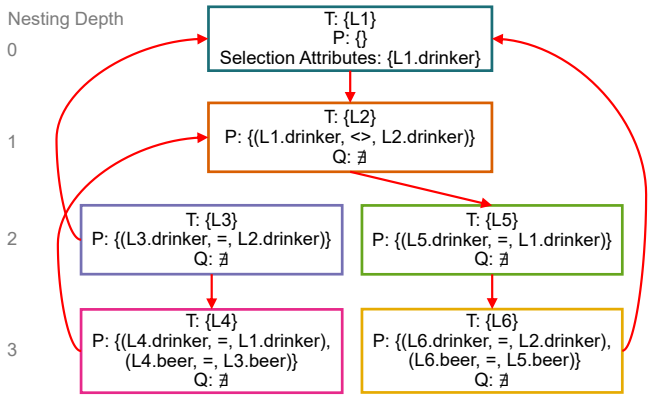

Figure 11: Intermediate representation of a QueryVis diagram superimposed on the logic tree structure illustrating with red edges the arrow direction between table groups that are connected via a join predicate. Notice how the arrows go from nodes with lower depth to nodes with greater depth if the difference is 1 and the opposite for depth difference greater than 1.

on the join predicates. Notice how the arrows go from nodes with lower depth to nodes with greater depth if the difference is 1 and the opposite for depth difference greater than 1. These arrow construction rules are crucial as they allow a more intuitive reading order over the QueryVis diagram as well as guarantee uniqueness as we shall demonstrate in Appendix B.

(5) Finally we create a SELECT table in the QueryVis diagram and append to it as rows all the attributes that are selected. The SELECT table is connected via undirected edges to the table attributes that are being selected. In our case we are selecting drinker from table L1. So we create a SELECT table with one row with value drinker and connect it to the drinker row of the L1 table.

Notice that the exact same procedure can be applied to construct the QueryVis diagram from the simplified LT Fig. 10b. The only differences are that tables L3 and L5 are surrounded by a double-lined bounding box and tables L4 and L6 are not bounded by a box. The QueryVis figures based on LTs from Figs. 10a and 10b are shown in Figs. 12a and 12b respectively.

### A.4 Rendering

Finally, we use GraphViz [32] to render the diagram.

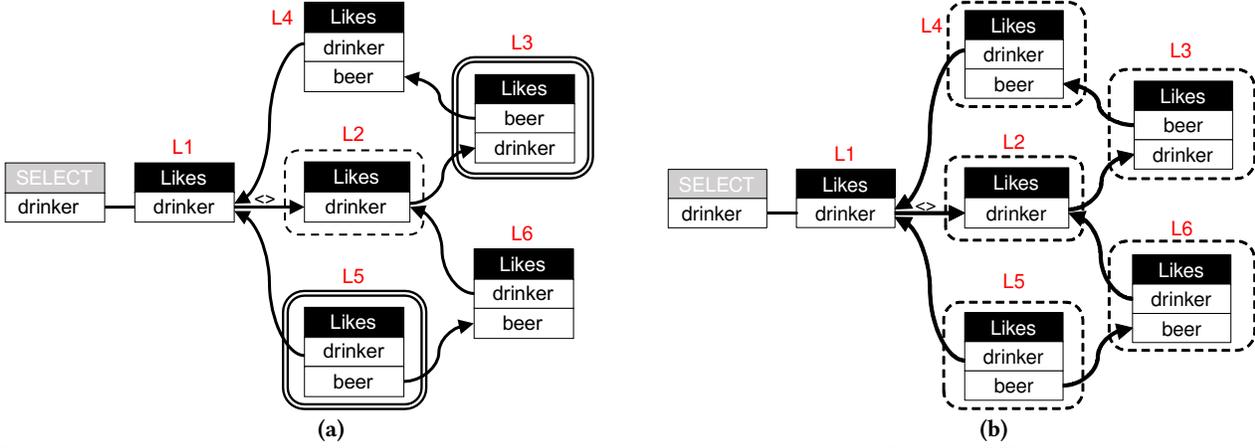

Figure 12: QueryVis diagrams based on the query from Fig. 1a. (a) was constructed from the LT shown in Fig. 10a and (b) was constructed from the LT shown in Fig. 12b.

## B  PROOF THAT VALID QUERYVIS DIAGRAMS ARE UNAMBIGUOUS

This section proves that we can always recover a unique logic tree (LT) from any valid QueryVis diagram (i.e., a diagram that was generated from a non-degenerate SQL query as defined in Section 5.1 of nesting up to depth 3). Recall that a LT is essentially an equivalent representation of the query in tuple relational calculus (TRC), which implies that every valid QueryVis diagram allow exactly one interpretation.

We present the proof into two parts. We first prove unambiguity for diagrams generated from path logic tree (i.e. queries where no query block has more than 1 directly nested query). Then then generalize to general queries.

### B.1  Path Logic Trees (Path-LTs)

We first consider LTs that are paths (i.e. no node in the LT has more than one child) of maximal depth 3. By construction, the query block[3] for each LT node is uniquely identified by a box in the QueryVis diagram. In the following discussion we view a QueryVis diagram as a directed connected graph $G$ whose nodes represent table groups. For non-ambiguity we therefore only need to show that one can unambiguously infer the nesting depth for each node in $G$.

Our proof demonstrates that there is a finite number of possible path-LTs and explore their QueryVis diagrams. Conceptually, each node in the QueryVis diagram has a hidden label denoting the depth of its corresponding node in the path-LT. We show that those labels can be recovered unambiguously, which implies that every valid QueryVis diagram has exactly one logical interpretation. For convenience, we show the hidden labels in the figures, but note that those labels would not be visible in an actual diagram.

---
[3] Recall a query block is a set of tables that are bounded by a quantifier. The contents of a query block correspond to a single node in an LT and an edge indicates a join condition between connected table groups.

First consider all possible types of edges in a QueryVis diagram of depth 3. There are 6 types of edges (marked as A-F in Fig. 13a where we label each node with the depth of the corresponding node in the path-LT), yet not all of the resulting $2^6 = 64$ patterns are valid. While we always know the node with depth 0, so our goal is to unambiguously determine the depth of the remaining 3 nodes labeled as (1, 2 and 3).

Our proof partitions the subset of 16 (= 8 + 4 + 4) valid patterns into 3 classes and proves unambiguity for each in turn. For that we denote by $\langle A \rangle$ the set of valid path-LT patterns for which the edge labeled $A$ (i.e., the edge between labeled node 1 and 2) is present. Similarly we use $\langle A, \bar{B} \rangle$ to denote the set of valid path-LT patterns for which the edge labeled $A$ is present and the edge labeled $B$ is absent. We now consider the 3 disjoint families of patterns $\langle A, B \rangle$, $\langle A, \bar{B} \rangle$ and $\langle \bar{A} \rangle$ illustrated in Figs. 13b to 13d and prove that we can always identify unambiguously the nesting depth for each node. It is easy to see that these 3 families cover all possible patterns (either edge A is present or not, and if then either B is present or not). Notice that a dotted edge implies that it is optional and does not have to be included in order for the path-LT to be a valid pattern.

- $\langle A, B \rangle$ (see Fig. 13b): Edge $D$ must be always present according to Property 5.2. We can always identify the root node, thus node 0 is in depth 0. Any outgoing edge from 0 needs to go to the node of depth 1 and thus since edge $A$ must be present we identify the node labeled 1 to be at depth 1. Similarly any outgoing edge from node 1 needs to go to the node of depth 2 and thus the node labeled 2 must be at depth 2. Finally any outgoing edge from node 2 either goes back to the root node

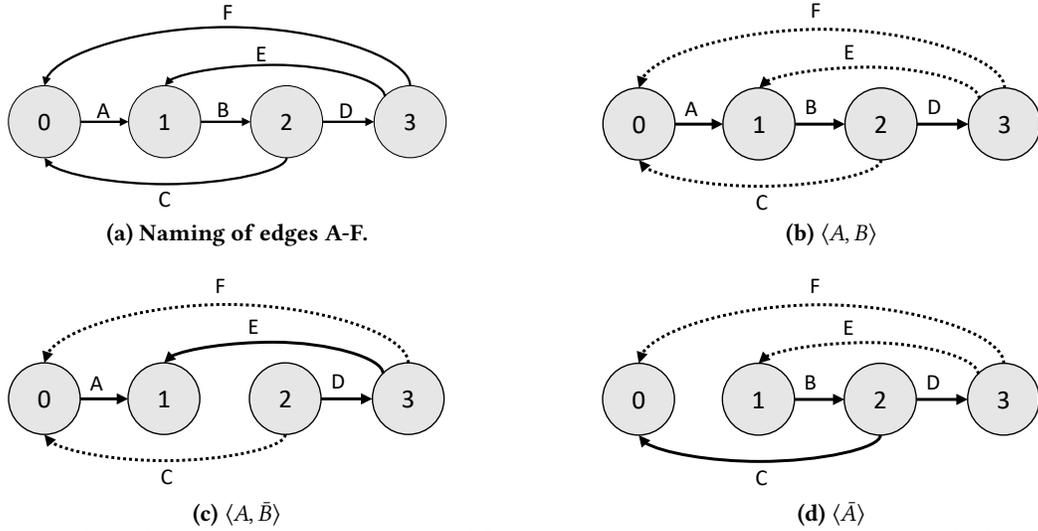

(a) Naming of edges A-F.
(b) $\langle A, B \rangle$
(c) $\langle A, \bar{B} \rangle$
(d) $\langle \bar{A} \rangle$

Figure 13: Figures of depth-3 patterns used in the proof for path logic trees in Appendix B.1: Dotted edges are optional. (a) Naming of the types of edges connecting different nesting depths. (b) The $\langle A, B \rangle$ pattern family where edges $A$ and $B$ must be present. (c) The $\langle A, \bar{B} \rangle$ pattern family where edges $A$ must be present and edge $B$ must not be present. (d) The $\langle \bar{A} \rangle$ pattern family where edge $A$ must not be present.

0 or to the node in depth 3. Since we already know which one is the root node we can identify that the node labeled 3 is at depth 3. Notice that edges labeled $C$, $E$ and $F$ are optional and including any combination of them will not affect the reasoning outlined above since edges $A$, $B$ and $D$ will always be present. We thus have $2^3 = 8$ possible patterns in that family.

- $\langle A, \bar{B} \rangle$ (see Fig. 13c): Edge $D$ must be always present according to Property 5.2. Since edge $B$ cannot be present, edge $E$ must be present in order to obey Property 5.2 (node at depth 1 must connect to its child node at depth 2 via its grandchild node in depth 3). We can always identify the root node, thus node 0 is in depth 0. Any outgoing edge from 0 needs to go to the node of depth 1 and thus since edge $A$ must be present we identify the node labeled 1 to be at depth 1. The node at depth 2 cannot have an incoming edge since we do not allow edge $B$, and since the node at depth 3 must have an incoming edge (i.e., $D$) from the node at depth 2 we can unambiguously identify that the node labeled 2 is in depth 2 and the node labeled 3 is in depth 3. Notice that edges labeled $C$ and $F$ are optional and including any combination of them will not affect the reasoning outlined above since edges $A$ and $D$ will always be present and edge $B$ is never present. We thus have $2^2 = 4$ possible patterns in that family.

- $\langle \bar{A} \rangle$ (see Fig. 13d): Edge $D$ must be always present according to Property 5.2. Since edge $A$ cannot be present, edge $B$ and $C$ must be present in order to obey Property 5.2 (the root node at depth 1 must connect to its child node at depth 1 via its grandchild node in depth 2). We can always identify the root node, thus node 0 is in depth 0. The node labeled 2 cannot be at depth 1 (because of edge $C$) and it also cannot be at depth 3 because if it was then the node labeled 3 would need to be in depth 1 which would violate Property 5.2. Therefore the node labeled 2 must be at depth 2. As a result the node labeled 1 must be at depth 1 due to edge $B$ and the node labeled 3 must be at depth 3. Notice that edges labeled $E$ and $F$ are optional and including any combination of them will not affect the reasoning outlined above since edges $B$, $D$ and $E$ will always be present and edge $A$ is never present. We thus have $2^2 = 4$ possible patterns in that family.

We have successfully shown that for all the depth-3 patterns from the 3 sets of patterns: $\langle A, B \rangle$, $\langle A, \bar{B} \rangle$ and $\langle \bar{A} \rangle$ we can unambiguously determine the nesting depth for each of the 4 nodes. Showing this ambiguity for path-LTs with nesting depth less than 3 is trivial as such patterns are sub-patterns from the ones shown in Fig. 13. Concretely, the problem of identifying all depth-2 pattern is solved by reducing from the problem of identifying all depth-3 patterns in the pattern family $\langle D, \bar{E}, \bar{F} \rangle$, thus edge $D$ present, and edges $E$ and $F$ absent. Those patterns are a strict subset of the patterns for which we have already proven unambiguity.

## B.2 General Logic Trees

We now generalize from path-LTs to arbitrary LTs of depth up to 3. While the number of patterns for paths up to depth 3 is finite, there are no restrictions on the number of branches a valid LT can have. This results in an unbounded number of unique QueryVis diagrams that can be constructed from a general LT. For our unambiguousness proof, we therefore cannot enumerate all these diagrams, but instead will devise decompositions to to transform a given QueryVis diagram produced for a general LT to a set of patterns we identified for path-LTs. This is done is by reducing a branching LT into a set of identifiable path-LTs that the branching LT is comprised of.

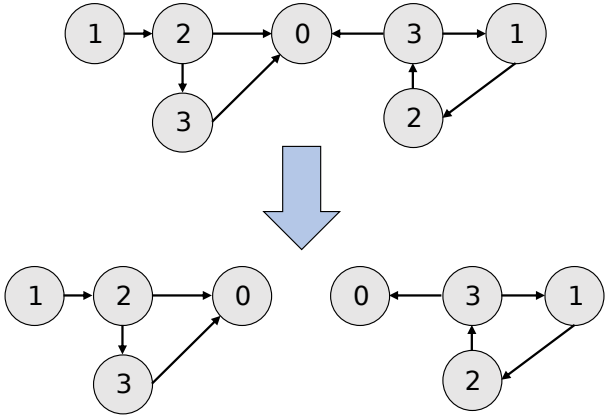

**Figure 14: An example QueryVis diagram decomposed using a depth 0 decomposition.** The root node 0 is removed creating two subgraphs and then added back to each subgraph separately with its appropriate edges. Notice that the resulting graphs are depth 3 path-LT patterns with edges {B, C, D, E} and {B, C, D, F} (see Fig. 13a).

Note that the LT has a single root whose corresponding node in the diagram is again uniquely identified by its connection from the SELECT box, but it may have more than one children. Similarly, nodes at depth 1 and 2 in the LT can also have multiple children. We next present a decomposition for each level. These decompositions are then applied in depth order, starting with the lowest.

*B.2.1 Depth-0 Decomposition.* The goal of this decomposition is to identify the subtrees of the root (depth 0) node. Given a directed graph $G$ representing a QueryVis diagram, we remove node 0 and all its connecting edges. Upon removal we are left with $k$ connected subgraphs (when ignoring edge directionality).

Those $k$ subgraphs correspond to the $k$ subtrees of the root. For each of the $k$ subgraphs we can attach the root node we removed with its corresponding edges connecting to the current subgraph and obtain a subtree of the root of the LT together with the root node.

Fig. 14 demonstrates this decomposition process. As shown in Fig. 14, after the depth 0 decomposition we are left with two patterns. Notice that Fig. 14 had a branch only at the root node (depth 0) so once we identified the subtrees we reduced our graph to one of the existing path-LT patterns, in this case depth-3 path-LT patterns with edges {B, C, D, E} and {B, C, D, F} (verify using the maximal depth-3 path-LT pattern in Fig. 13a). This won't be true in general as we can also have arbitrary branching starting at nodes of depth 1 and 2, so we will need further decompositions for depth 1 and 2 as we discuss below.

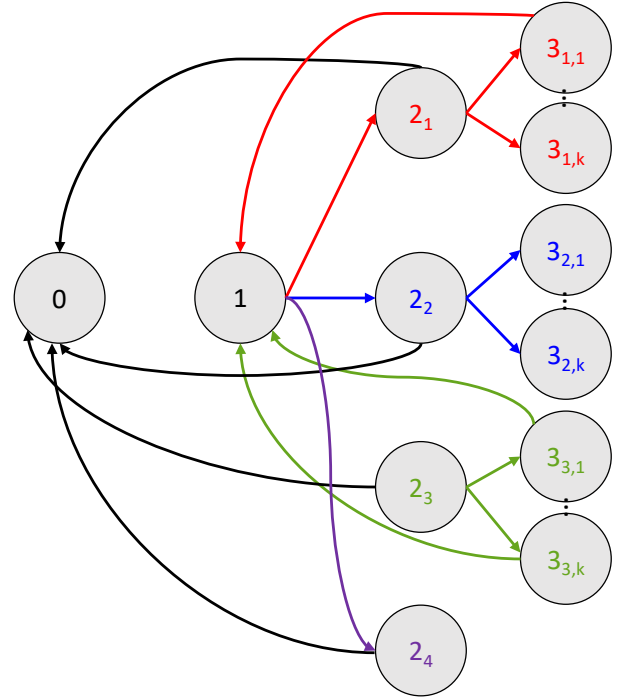

**Figure 15: All 4 possible patterns/ways (shown in color) a depth 1 node can be connected in a subgraph produced from a depth 0 decomposition when the root node is not directly connected to the node of depth 1.**

*B.2.2 Depth 1 Decomposition.* The goal of the depth 1 decomposition is to identify the node(s) that are in depth 1 and as a result we can identify the subtrees of the depth 1 nodes. The procedure for the depth 1 decomposition is more involved than that of the 0 depth decomposition as there are more possibilities to reason over. Nevertheless, the possibilities are finite and we track them on a case by case basis.

As we discussed, before performing the depth 1 decomposition we perform a depth 0 decomposition which gives us the $k$ subgraphs that correspond to the $k$ subtrees of the root node in the LT. We also attach the actual root node back into each subgraph, just like we did in the bottom row of Fig. 14. For each of those subgraph there are two possibilities:

**(1) Root node has an outgoing edge.** An outgoing edge from the root node indicates that the node it connects to must be its child which is the target node we want to identify at depth 1. Since we are reasoning over a subgraph produced by the depth 0 decomposition it will contain only one node at depth 1. So if the root node has an outgoing edge we can immediately identify the node at depth 1.

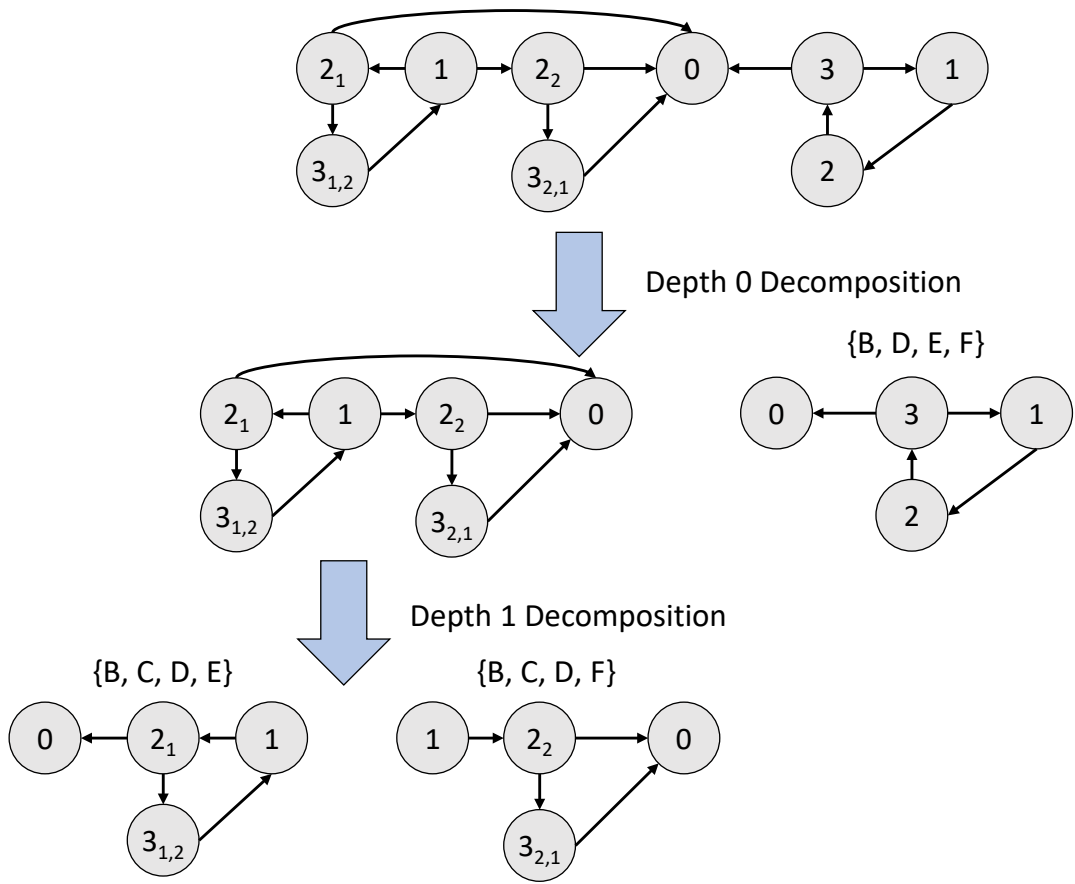

Figure 16: An example QueryVis diagram decomposed using both depth 0 and 1 decompositions. Notice how after decomposing the graph in depth 0 one of the subgraphs is reduced to a pattern we know, namely the depth-3 path-LT pattern with edges {B, D, E, F}. Next we identify on the remaining subgraph that the node labeled 1 is at depth 1 and thus after we perform the depth 1 decomposition we reduce our subgraph into two more patterns that we can identify, namely the depth-3 path-LT patterns with edges {B, C, D, E} and {B, C, D, F}. The 2 subgraphs produced after the depth 1 decomposition were found by first removing nodes 0 and 1, then identifying the resulting subgraphs and then for each one attaching back nodes 0 and 1.

**(2) Root node has no outgoing edge.** This case is more complicated as we cannot directly infer the node in the subgraph that is at depth 1. There are 4 possible ways/patterns (all of them can occur simultaneously with arbitrarily many duplications) our node in in depth 1 could be connected in the subgraph.

(1) The node in depth 1 is connected to a node in depth 2 which in turn has some or all of its children connect back to the node in depth 1 (branch $1 \rightarrow 2_1$ shown in red color in Fig. 15).
(2) The node in depth 1 is connected to a node in depth 2 which in turn has none of its children connect back to the node in depth 1 (branch $1 \rightarrow 2_2$ shown in blue color in Fig. 15).
(3) The node in depth 1 connects to all the children nodes of a node in depth 2 but is not connected to that depth 2 node (subtree starting at $2_3$ shown in green color in Fig. 15).
(4) The node in depth 1 connects to a depth 2 node that has no children (subtree starting at $2_4$ shown in purple color in Fig. 15.

Given the above possible set of 4 patterns in any such subgraph we can now unambiguously determine which node is at depth 1. Notice that the nodes labeled as $2_1, 2_2, 2_3, 2_4$ in Fig. 15 cannot be at depth 1 because if they did the direction of the edge with the root node would be the other way around. Notice that all the nodes in depth 2 would necessarily have to be connected to the root node in order to obey Property 5.2 if the root node is not directly connected to the node in depth 1. Thus the candidate nodes for our our node in depth 1 are $1, 3_{1,1}, \ldots, 3_{1,k}, 3_{2,1}, \ldots, 3_{2,k}, 3_{3,1}, \ldots, 3_{3,k}$. To determine which which one of our candidates is at depth 1 we perform the following procedure:

Remove the root node and its accompanying edges. We are now left with a subgraph $G$, a graph as shown in Fig. 15 without the root node and the black colored edges. For each candidate node, remove it from the subgraph $G$ together with its accompanying edges and observe if the following on the remaining subgraph $G'$ is true.

- If $G'$ is composed by more than 1 connected subgraphs then this means that the graph $G$ had multiple nodes at depth 2 and our candidate node is the node at depth 1. In that case we are done and have identified the node at depth 1.

If the above statement was never true for any candidate then that means the node at depth 1 has only 1 child node. In that case there are two possibilities:

- G was already an already existing pattern (i.e., a path-LT), so we are done as we can always identify the depth of each node in a path-LT as we have shown in Appendix B.1.
- G is not a path-LT which implies that the node in depth 2 has multiple children nodes. We can easily identify the node with depth 2 as it will be the node with the greatest out degree (remember that there can only be on depth 2 node in this scenario and thus the out-degree of the node in depth 1 is at most 1). Thus the node at depth 1 is the node that connects directly to the node in depth 2 or indirectly via all the children of the node in depth 2.

In any case, as we have outlined above we will always be able to identify the node in depth 1 from graph $G$.

Once we have identified the node in depth 1, we can identify all the $l$ subtrees starting from a node of depth 1 just like we did with the depth 0 decomposition in Appendix B.2.1. For each of those $l$ subtrees we attach back again the depth 0 node and the depth 1 node we identified to create $l$ subgraphs that are the output of the depth 1 decomposition. Those subgraphs are either one of our existing patterns and we are done or if that's not the case then it means that the node in depth 2 has multiple children and thus we need to apply a depth 2 decomposition on the subgraph.

Fig. 16 demonstrates the depth 1 decomposition with a QueryVis diagram that was produced by a LT where the root can have many children nodes and a depth 1 node can also have many children nodes a well.

B.2.3 *Depth 2 Decomposition.* Now to completely generalize to any possible branching in a LT we consider all the possible arbitrary number of children a node in depth 2 can have. Recall that in the depth 1 decomposition Appendix B.2.2, it is possible that our output subgraph is still not one of the patterns because of multiple children from a depth 2 node. We can resolve this issue by applying a depth 2 decomposition.

The goal of the depth 2 decomposition is to identify the node of depth 2 in a given subgraph produced after a depth 1 decomposition. Any output subgraph of a depth 1 decomposition has only one node of depth 0, 1 and 2. If a depth 2 node has many children then it must connect to all of them via outgoing edges in order to obey Property 5.2. Therefore if we remove the depth 0 node and its edges from the subgraph, in the remainder subgraph the node with the highest out-degree in the subgraph is the node in of depth 2. This is because the maximum out-degree of the depth 1 node is 1, the maximal out-degree of a depth 3 node is 1 (connects to the depth 1 node), so only the node of depth 2 can have an out-degree greater than 1. If the out-degree of every node is 1 then the node in depth 2 has only one child and a depth 2 decomposition is not necessary as the subgraph would be one of our already known patterns in Appendix B.1.

Fig. 17 shows a complete example of a QueryVis diagram that needs decomposition at depths 0, 1, and 2.

With the following set of 3 decompositions we have shown that for any QueryVis diagram that was produced by an arbitrarily branching LT, we can always decompose the diagram into one of the finite patterns we identified in Appendix B.1 for which we know how to construct its corresponding LT.

## C DETAILS ON USER STUDY
## C.1 Reproducibility and OSF

In our original submission to SIGMOD 2020, we had reported the results of a user study with $n = 102$ participants on Amazon Turk. Due to questions by the reviewers on reproducibility, and based on our experiences with the first user study and additional careful reading of related work, we decided to run a completely new study. We preregistered this study on the Open Science Framework (OSF) at `osf.io/vy2x3` on January 24th, 2020 — the day before we started running the study on Amazon Mechanical Turk (AMT). That way, our intended analyses are immutable and timestamped, including the code to analyze the data. The full study materials includes all details of our study (including the description, the question stimuli, the answer choices, the data collected, and the Python Jupyter Notebook used to analyze the data and create our figures) are available on OSF at `osf.io/mycr2` which allows reproducible results. We also posted all the intended analysis of our user study on OSF before actually running it. That allows reproducibility and transparency on all assumptions made. Due to the time limitation for resubmission we got only $n = 42$ legitimate participants vs.

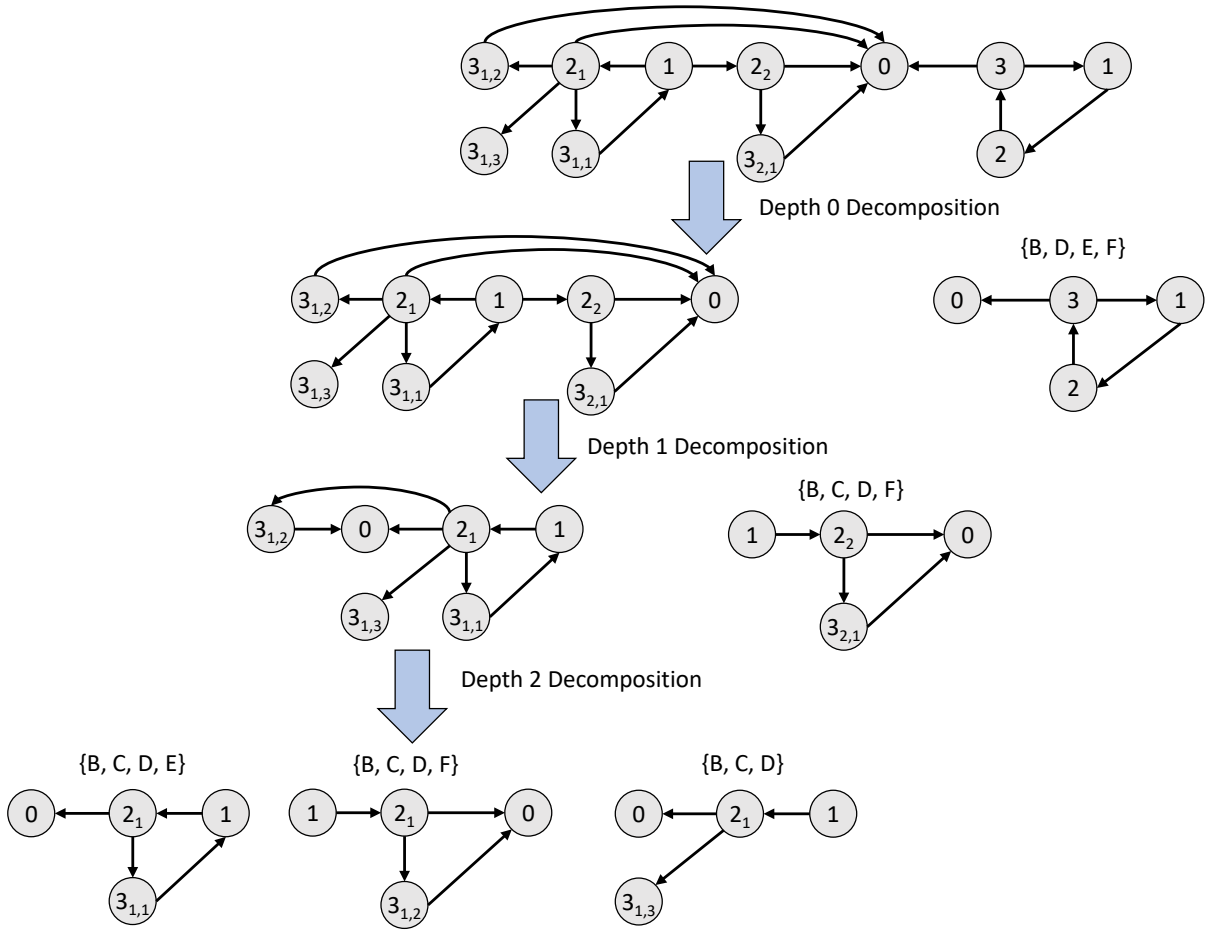

Figure 17: An example QueryVis diagram decomposed using all depth 0, 1, and 2 decompositions. This example is very similar to the one illustrated in Fig. 16 with the only difference that the node labeled $2_1$ has 3 child nodes namely, $3_{1,1}, 3_{1,2}$ and $3_{1,3}$. Notice during the depth 2 decomposition step we identify that node $2_1$ has the highest out-degree and thus is the node in depth 2. As it has 3 children nodes we have 3 subgraphs as shown in the bottom row. All 3 subgraphs fall into one of the patterns identified in Appendix B.1.

$n = 102$ for the original study. The tutorial for our study is listed in Appendix E, the 12 questions we used in Appendix F.

## C.2 Arguments for a within-subjects study

We elaborate here on our justifications for using a *within-subjects study* — in which every participant sees the same fraction of queries in each of the 3 modes (SQL, QueryVis, Both) — versus using a *between-subjects study* — in which every participant would only see one of the modes. As a consequence of this choice and our decision to use a Latin Square design (see Section 6.1), the number of SQL questions in our test needed to be a multiple of 3.

*C.2.1 Added statistical power in a smaller study.* Pilot testing indicated high variability in time and error between participants. Thus we chose a within-subjects design to minimize the effect of individual differences and increase power [55]. Within-subject designs are widely known to provide more power with same number of participants, thus allowing smaller studies [40].

*C.2.2 Better external validity.* We expect real-world users of QV would be using it alongside SQL or will alternate between views depending on their task. E.g., they may copy and edit SQL while search for a query using QV. Moreover, we expect that a user will view and interpret multiple queries before selecting the one of interest or writing their own. Thus, we argue that our study design has a better external (ecological) validity than a between-subjects design would [40].

*C.2.3 Avoiding Complications.*

**(1) Avoiding unwanted practice effects.** We are interested in the treatment effect of QV on a practiced skill (reading a query). If we were only showing three questions, one in each mode, a within-subjects study may not have been as appropriate. However, we are showing each mode multiple times separated by practice in other modes, thus there is less need to provide extensive practice ahead of time. Moreover, we help avoid unwanted practice effects using Latin square counterbalancing on the order modes are presented [40, 55].

**(2) No predicted sensitization effects.** We believe sensitization effects to be unavoidable when presenting a novel visualization as a replacement or augment for standard practice (SQL). As extensive SQL knowledge is required in order to interpret any of our modes (all based on SQL), making a comparison to SQL is only natural. Thus there is no reason to avoid a within-subjects study because it might highlight the distinction.

**(3) Little predicted carry-over effect.** We do not expect that viewing a question in one mode will meaningfully affect participant performance when they subsequently use a separate mode for a new question. Regardless, our counterbalancing approach helps alleviate any problematic effect [40, 55].

### C.3 Choice of queries and answers

We describe here the process by which we designed the questions used in our user study.

**Types of SQL queries.** At the early stage of our question design, we suspected there is a correlation between the features of SQL queries and the difficulty for users to understand them. This led us to examine the common types of SQL errors analyzed by past work [4, 11]. Based on the error rate and how the authors categorized the queries used in their study (especially [4]) we found that nested SQL queries, especially with correlated subqueries and/or self-joins, tend to form the most challenging questions and result in a lower success rate for test takers than other types of features (e.g., simple conjunctive and natural joins). This observation is also in line with work dating back to the early days of SQL that suggested that users find quantification to be particularly difficult [66, 67]. The entirety of the discussion in these previous studies led us to identify 4 meaningful categories of SQL questions:

(1) conjunctive queries without self-joins,
(2) conjunctive queries with self-joins,
(3) nested queries, and
(4) queries with groupings.

For each category, we created 3 queries: one simple, one medium, and one complex query. The query complexity was designated based on the number of joins and number of table aliases referenced in the query. This led to 12 (= 3 × 4) total queries, which are listed in Appendix F.

**Extension to GROUP BY queries.** In our study, we also tested a novel extension of our covered SQL fragment to groupings, which is not focus of our main paper. Thus 3 of the 12 questions on queries with groupings, the remaining 9 questions focus on the SQL fragment without grouping. We report the results on the 9 questions without groupings in the sections of the paper Section 6, and give details on the full 12 questions here in Appendix C.5.

**4 meaningful answer choices.** We early decided to use multiple-choice questions (MCQs) as they allow *unambiguous grading* (answers by participants one of several predetermined choices) and *scalable testing* (adding more participants to a study does not increase the amount of work). Yet MCQs need more to be designed correctly. When designing our MCQs, we followed best practices [85] that advise that no choice should be easy to eliminate without actually understanding SQL. For example, a simple selection criterion like *boats that are green* vs. *boats that are red* could be discerned without even knowing SQL. But deciding whether a query finds *sailors that reserve only red boats* vs. *sailors that reserve all red boats* is not trivial and requires understanding the join logic. Our test was formulated as a multiple-choice test, we thus we designed all 4 of the choices to read very similar to each other. This was so that a participant with little knowledge of SQL would be incapable of eliminating any of the 4 choices. The doctrine of our MCQ is to make incorrect choices misleading but *not ambiguous in natural language*. Thus for all questions the differences between the 4 choices were all at the logical level and no answer choice could be eliminated by a quick glance over the query or diagram (i.e., no choice references spurious tables or values that aren't present in the query). At the same time, they were also not ambiguous in the natural language representation. Thus the questions focused on difficulty of SQL understanding instead of interpretations of ambiguous wording. The full list of queries and answer choices are listed in Appendix F and on OSF at osf.io/mycr2

**Database schema.** We decided to use one schema for all queries, including the queries in the qualification exam. This allowed us to separate *schema understanding* from *query understanding*. We decided on using the Chinook database [20] which is widely used across various classes teaching SQL and rich enough to model real business problems.

### C.4 Qualified and legitimate participants

Participants in our study had to get a 4/6 questions correct in the qualification test. We had 710 AMT workers attempt our qualification SQL test, and only 114 passing the qualification.

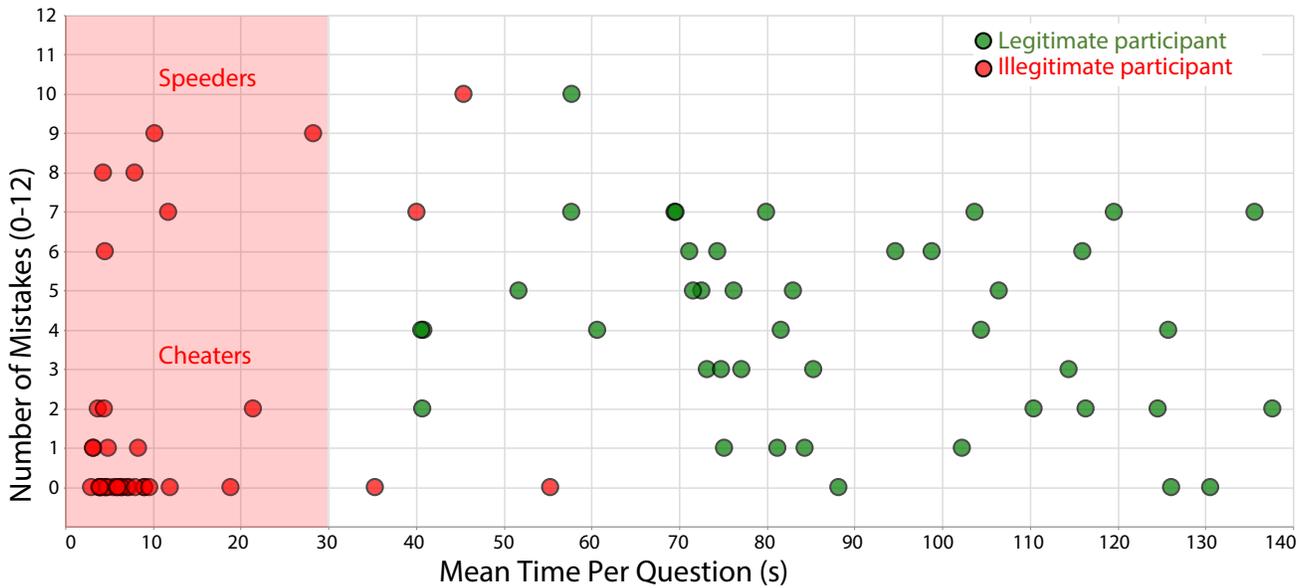

Figure 18: Scatter plot of the mean time per question (x-axis) versus the number of mistakes (y-axis) for all 80 of our participants. We have colored in red the 38 participants that were deemed illegitimate (either speeders or cheaters). Notice the cheaters can be found in the bottom left corner of the scatter plot with zero or almost zero mistakes and extremely short mean time per question. The speeders are scatter at the top left of the scatter plot where they answered fast and randomly thus a lot of mistakes. We classified users that tool needed than 30 seconds per question as speeders/cheaters. We also identified 4 participants with a mean time greater than 30 seconds per question as speeders/cheaters (2 speeders and 2 cheaters). The 2 speeders did the test normally up to a point and then speeded through the last few questions as they most likely gave up. The 2 cheaters made zero mistakes and had extremely short completion times for all questions except 1 question which explains the bump in their time per question. We are confident that the 42 participants marked in green were all legitimate and carefully spend their time in each question.

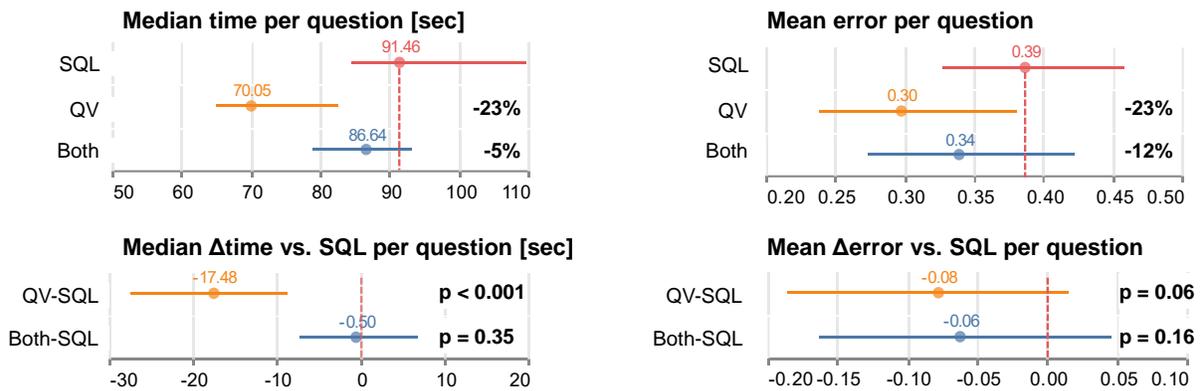

Figure 19: Median time per question and mean error rates for all 12 questions (including the 3 Group-By questions) for each of the 3 conditions with their respective 95% BCa confidence intervals (CIs) that show the range of plausible values for the median and means respectively. Participants were on average faster using QV than SQL (-23%, $p < 0.001$), and slightly faster using Both than SQL (-5%, $p = 0.35$). The differences in error rate were also noticeable with participants making less errors using QV than SQL (-23%, $p = 0.06$) and less errors with Both than SQL (-12%, $p = 0.16$). Note that the $p$-values are based on within-subjects statistical tests, thus confidence intervals of the mean could overlap substantially without indicating a high $p$-value [29]. The $p$-value for a given hypothesis tests an individual's performance in the QV or Both condition vs. their performance in the SQL condition. The $p$-values are listed along with the respective percentage difference of the condition's values vs. their counterpart SQL value. However, the confidence intervals capture the median/mean of a given condition for all participants, neglecting analysis at a within-participant level. The bottom row shows the per-participant differences of QV or Both condition vs. their performance in the SQL condition.

Among those successfully completing our qualification test, 80 participants started the study.

However upon further inspection we identified a large proportion ($n = 38$) of participants were either *speeders* (participants that went through questions very fast in hopes of answering the question correctly at random) or *cheaters* (participants that somehow received the correct answers from others and managed to get no or very few mistakes with extremely low completion time). In order to identify such problematic participants we plotted the mean time per question versus the number of mistakes they made (out of 12) for our 80 participants in a scatter plot shown in Fig. 18. As seen from the figure, the cheaters are clustered at the bottom left of the figure (low mean time per question and very few mistakes). The speeders are at the top left (low mean time but numerous mistakes). By observing the distribution of completion times and considering the difficulty of each question we identified 30 seconds per question to be a good cutoff point distinguishing between legitimate participants and speeders/cheaters. Upon further examination we also identified 4 more participants (shown in red a the right of the 30 second cutoff line Fig. 18) that were speeders or cheaters. The two additional speeders had high mean time per question because they gave up mid-test (i.e., speeded over a portion of the test). The two additional cheaters waited very long on a specific question and then answered the remaining ones very fast and correctly. In all our analyses we removed the participants whom we deemed illegitimate and refer to the remaining 42 participants as *legitimate*.

## C.5 Study Results

We perform our analyses on both the 9 questions without GROUP BY (as presented in Section 6), as well as on all 12 questions (i.e., including the 3 GROUP BY questions). Appendix F lists all 12 questions.

The study results on the 9 questions are shown in Fig. 7 and Fig. 20. The study results on all 12 questions (i.e., including the 3 Group-By questions) are shown in Fig. 19 and Fig. 21. Figs. 20 and 21 are particularly useful to look at as they show the distribution of the differences in time and error between QV and SQL on a per participant basis. Figures 20a and 21a show that 71% and 76% of our participants were faster with QV than SQL in our 9 and 12 questions analyses. Figures 20b and 21b show that more participants tend to make fewer mistakes when using QV instead of SQL (36% vs. 26%, and 40% vs. 29%).

Overall, we notice that there do not appear to be meaningful differences in our results depending on whether we include the 3 Group-By questions in our analysis. Therefore, we are confident that QueryVis works on grouping as well without groupings without affecting user performance.

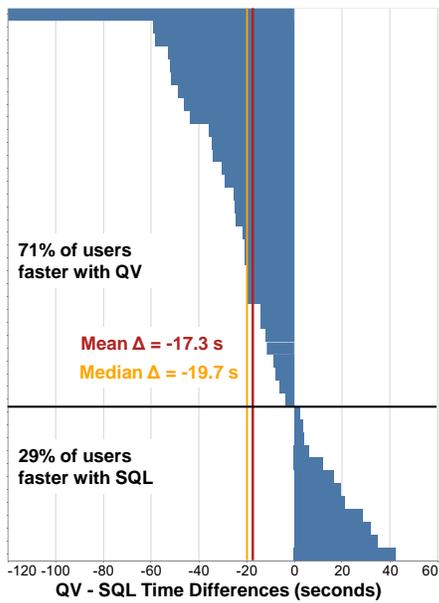
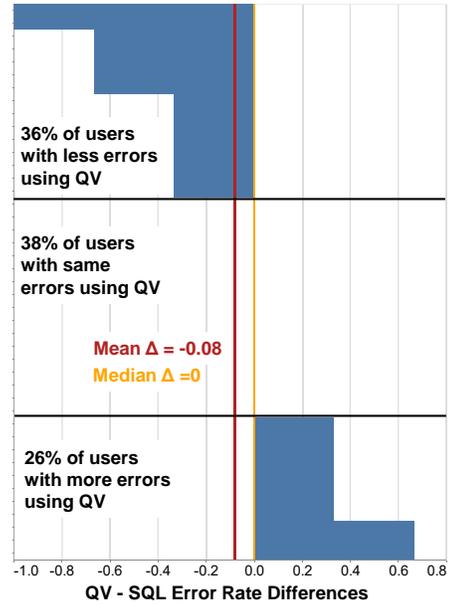

(a)      (b)

Figure 20: QueryVis – SQL time and error differences for each participant using 9 questions (Group-By questions excluded).

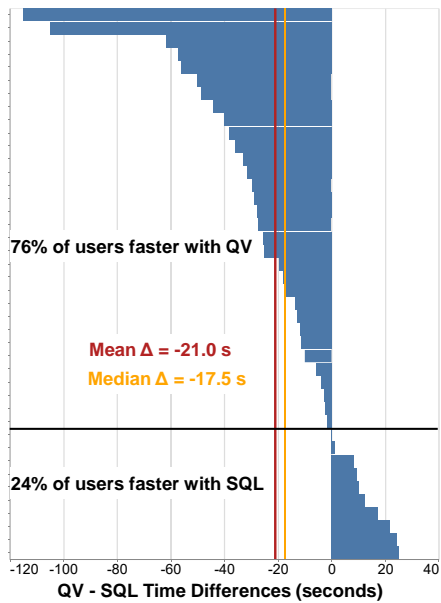
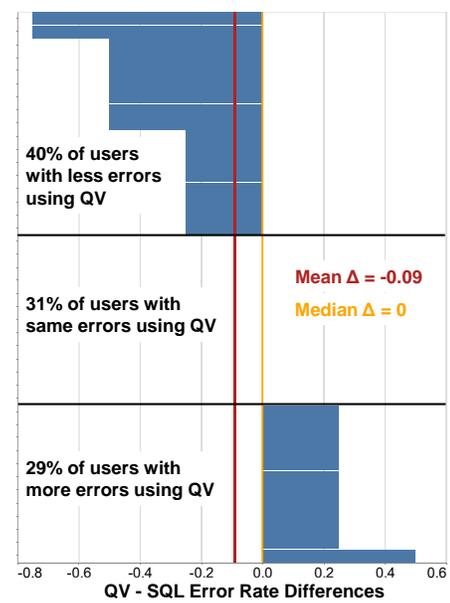

(a)      (b)

Figure 21: QueryVis – SQL time and error differences for each participant using all 12 questions (including Group-By questions).

# D STUDY DETAILS: 6 QUALIFICATION QUESTIONS

We include here the 6 qualification questions. Participants had to get 4/6 questions correct in order to pass the SQL qualification.

## Qualification Question #1

```
SELECT  P.PlaylistId, P.Name
FROM    Playlist P, PlaylistTrack PT, Track T,  Album AL, Artist A
WHERE   P.PlaylistId = PT.PlaylistId
AND     PT.TrackId = T.TrackId
AND     T.AlbumId = AL. AlbumId
AND     AL.ArtistId = A.ArtistId
AND     A.Name = 'AC/DC';
```

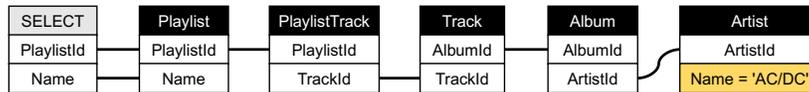

A. Find playlists that have all tracks from all albums by artists with the name 'AC/DC'.
B. Find playlists that have all tracks from an album by an artist with the name 'AC/DC'.
C. Find playlists that only have tracks from albums by artists with the name 'AC/DC'.
D. Find playlists that have at least one track from an album by an artist with the name 'AC/DC'.

## Qualification Question #2

```
SELECT  C.CustomerId, C.FirstName, C.LastName
FROM    Customer C, Invoice I,
        InvoiceLine IL1, InvoiceLine IL2,
        Track T1, Track T2
WHERE   C.CustomerId = I.CustomerId
AND     I.InvoiceId = IL1.InvoiceId
AND     I.InvoiceId = IL2.InvoiceId
AND     IL1.TrackId = T1.TrackId
AND     IL2.TrackId = T2.TrackId
AND     T1.GenreId <> T2.GenreId;
```

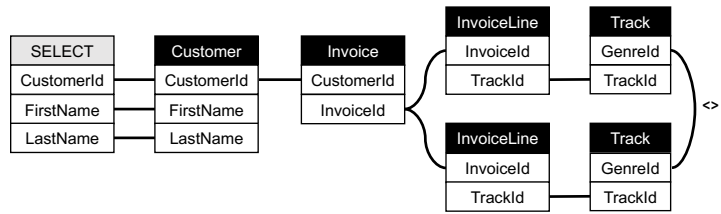

A. Find customers who have at least two invoices and for each invoice there are at least two tracks of different genres.
C. Find customers who have at least two invoices with tracks of different genres.
B. Find customers who have an invoice with at least two tracks of different genres.
D. Find customers who have an invoice with only two tracks that are of different genres.

# Qualification Question #3

```
SELECT      P.PlaylistId, G.Name, COUNT(T.TrackId)
FROM        Playlist P, PlaylistTrack PT, Track T, Genre G
WHERE       P.PlaylistId = PT.PlaylistId
AND         PT.TrackId = T.TrackId
AND         T.GenreId = G.GenreId
GROUP BY    P.PlaylistId, G.Name;
```

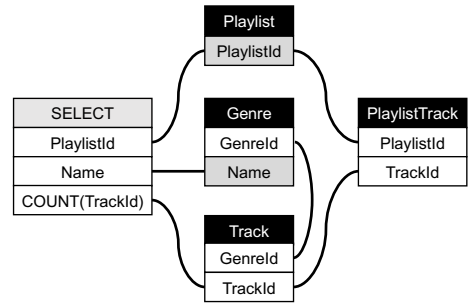

A. For each playlist, find the number of tracks per genre.
B. For each genre, find the number of tracks in the genre.
C. For each playlist find the number of tracks in the playlist.
D. For each playlist and genre, find the number of tracks in each playlist.

# Qualification Question #4

```
SELECT   A.ArtistId, A.Name
FROM     Artist A
WHERE    NOT EXISTS
         (SELECT  *
          FROM    Album AL
          WHERE   AL.ArtistId = A.ArtistId
          AND     NOT EXISTS
                  (SELECT  *
                   FROM    Track T, MediaType MT
                   WHERE   AL.AlbumId = T.AlbumId
                   AND     T.MediaTypeId = MT.MediaTypeId
                   AND     MT.Name = 'ACC audio file')
         );
```

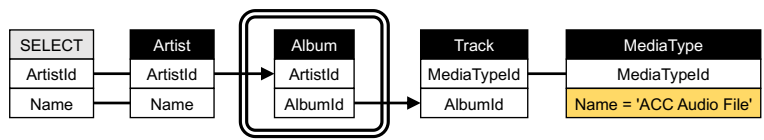

A. Find artists where all tracks in all their albums are available in 'ACC audio file' type.
B. Find artists where all their albums have a track that is available in 'ACC audio file' type.
C. Find artists where none of their albums have a track that is available in 'ACC audio file' type.
D. Find artists where none of their albums have all their tracks available in 'ACC audio file' type.

# Qualification Question #5

```
SELECT      C1.CustomerId, C1.FirstName, C1.LastName
FROM        Customer C1, Invoice I1, InvoiceLine IL1,
            Track T1, Album AL1, Artist A1
WHERE       C1.customerId = I1.CustomerId
AND         I1.InvoiceId = IL1.InvoiceId
AND         IL1.TrackId = T1.TrackId
AND         T1.AlbumId = AL1.AlbumId
AND         AL1.ArtistId = A1.ArtistId
AND         A1.Name = 'AC/DC'
AND         NOT EXISTS
            (SELECT *
             FROM    Customer C2, Invoice I2, InvoiceLine IL2,
                     Track T2, Album AL2, Artist A2
             WHERE   C2.CustomerId <> C1.CustomerId
             AND     C1.City = C2.City
             AND     C2.CustomerId = I2.CustomerId
             AND     I2.InvoiceId = IL2.InvoiceId
             AND     IL2.TrackId = T2.TrackId
             AND     T2.AlbumId = AL2.AlbumId
             AND     AL2.ArtistId = A2.ArtistId
             AND     A2.Name = 'AC/DC');
```

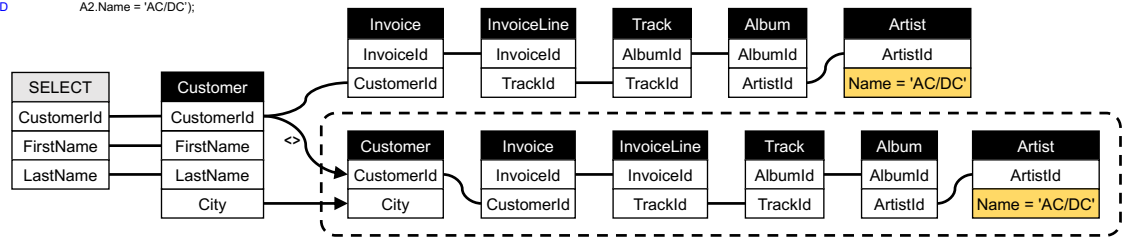

A. Find customers who were not the only ones in their city to buy every track from an album by an artist with the name 'AC/DC'.
B. Find customers who were the only ones in their city to buy every track from an album by an artist with the name 'AC/DC'.
C. Find customers who were not the only ones in their city to buy a track from an album by an artist with the name 'AC/DC'.
D. Find customers who were the only ones in their city to buy a track from an album by an artist with the name 'AC/DC'.

# Qualification Question #6

```
SELECT      E1.EmployeeId, COUNT(C.CustomerId), AVG(I.Total)
FROM        Employee E1, Employee E2, Customer C, Invoice I
WHERE       E1.ReportsTo = E2.EmployeeId
AND         E1.Country <> E2.Country
AND         E1.EmployeeId = C.SupportRepId
AND         E1.Country = C.Country
AND         C.CustomerId = I.CustomerId
GROUP BY    E1.EmployeeId;
```

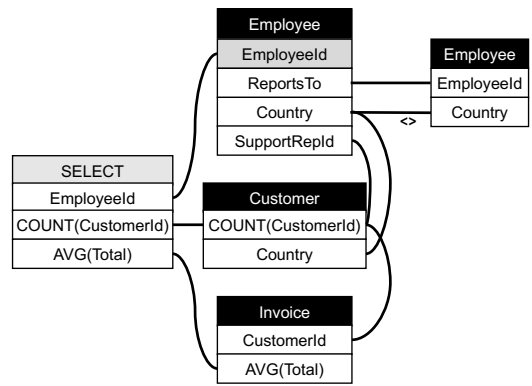

New Choices:
A. For each employee that reports to an employee in another country, find the number of customers the former employee services in a different country than theirs and the average invoice total of those customers.
B. For each employee that reports to an employee in another country, find the number of customers the former employee services in their country and the average invoice total of those customers.
C. For each employee that reports to an employee in another country, find the number of customers the latter employee services in a different country than theirs and the average invoice total of those customers.
D. For each employee that reports to an employee in another country, find the number of customers the latter employee services in their country and the average invoice total of those customers.

# E STUDY DETAILS: 10-PAGE TUTORIAL

We include here the 10-page tutorial that our study participants went through before answering the test questions.

**Visual diagrams for interpreting existing SQL queries?**

- The goal of the following 9 pages is to provide you with a quick introduction to our study setup on AMT, an overview of the database schema used for all the 12 queries in the study, and a tutorial on how to read the visual diagrams.

- Use the buttons below or the keyboard's left and right keys to navigate through the tutorial.

Page: 1/10

All questions during the test will be using the relational schema of a music publisher's digital media store, including tables for artists, albums, media tracks, invoices and customers.

On the right, primary keys are underlined, and foreign keys point to the primary key they refer to.

- A track is from one album.
- Each album is by one artist.
- A track has one media type (mp3, AAC, etc.)
- A track has one genre (pop, rock, rap, etc.)
- A purchased track has an invoice line.
- An invoice is created each time a customer makes a purchase.
- An invoice has one or more invoice lines, one for each track purchased.
- The invoice line "quantity" attribute records how many copies of the same track were purchased in one invoice. Thus customers can purchase the same track multiple times!
- Attribute "Milliseconds" records the duration/length of a track.
- Each customer has an employee support representative.
- Each employee reports to another employee.

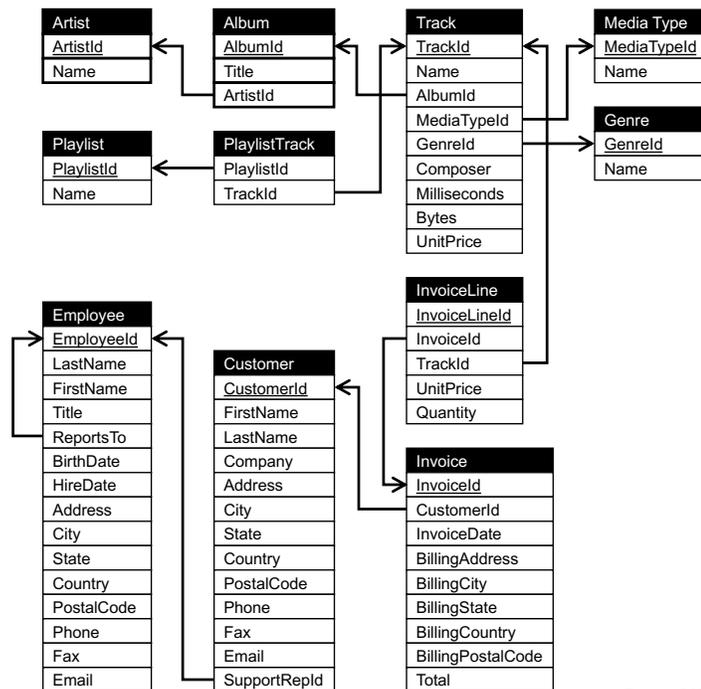

Page: 2/10

# Basic Conjunctive Query

The visual diagram for a query is read as follows:

*(1) Select the attribute TrackId …*  *(2)… from the table Track…*  *(3)… where unit price is greater than 2.*

```
SELECT    T.TrackId
FROM      Track T
WHERE     T.UnitPrice > 2;
```

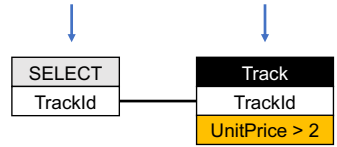

- *The diagrams show predicates like "T.UnitPrice > 2" with a yellow background.*

Interpretation:
Find TrackId of Tracks whose UnitPrice is greater than 2.

---

# Basic Conjunctive Query with Implicit and Explicit Join Syntax

If you are unfamiliar with implicit join, here's an example:

Implicit Joins:

```
SELECT    T.TrackId
FROM      Track T, PlaylistTrack PT, Playlist P, Genre G
WHERE     T.GenreId = G.GenreId
AND       T.TrackId = PT.TrackId
AND       PT.PlaylistId = P.PlaylistId;
```

Explicit Joins:

```
SELECT    T.TrackId
FROM      Track T
JOIN      Genre G              ON  T.GenreId = G.GenreId
JOIN      PlaylistTrack PT     ON  T.TrackId = PT.TrackId
JOIN      Playlist P           ON  PT.PlaylistId = P.PlaylistId;
```

- *Two queries above are equivalent.*
- *In this assignment, all SQL joins are inner and will be written as implicit joins.*

Interpretation:
"Find the TrackId of Tracks that are in some Playlist and belong to some Genres."

# Basic Query With Joins

```
SELECT    T.TrackId
FROM      Track T, PlaylistTrack PT, Playlist P, Genre G
WHERE     T.GenreId = G.GenreId
AND       T.TrackId = PT.TrackId
AND       PT.PlaylistId = P.PlaylistId
AND       G.Name <> P.Name;
```

The visual diagram for a query is read as follows:

*(1) Select the attribute TrackId …*
*(2)… from the table Track where the Track …*
*(3)… is in a playlist…*
*(4)… whose name is different from the track's genre.*

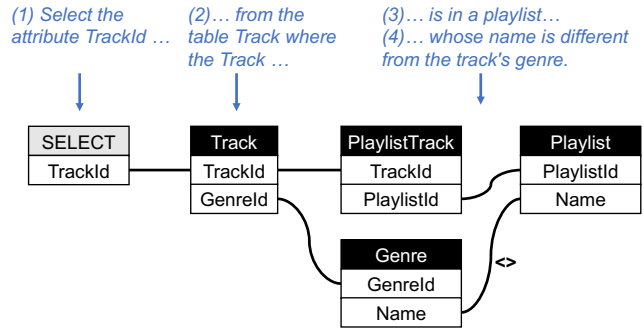

- *Cross-table joins are represented by a line connecting the joining attributes from the two tables.*
- *An unlabeled line represent an equijoin (=).*
- *A labeled line represents a join applying the logic operator of its label. In the example above, the line between G.Name and P.Name is labeled with <>, indicating that we join using the 'not equals' operator.*

Interpretation:
"Find the TrackId of Tracks that are in some Playlist whose name is different from the Genre of the Track."

# Group By Queries with Aggregates

```
SELECT     T.TrackId, SUM(IL.Quantity)
FROM       InvoiceLine IL, Invoice I
WHERE      IL.InvoiceId = I.InvoiceId
AND        I.CustomerId = 123
GROUP BY   T.TrackId;
```

The visual diagram for a query is read as follows:

*(1) Select the attributes TrackId and SUM(Quantity)…*
*(2)… from table InvoiceLine **grouped by** TrackId and **summed up** Quantities …*
*(3)… for invoices from the customer with ID 123.*

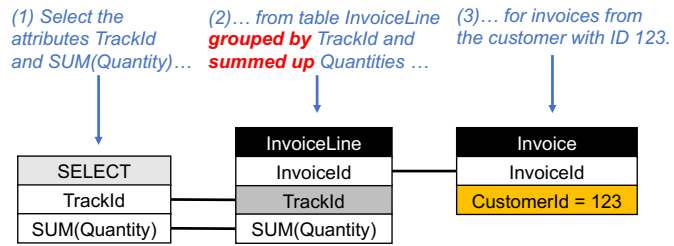

- *The attribute that we "**Group By**" is highlighted with a grey background.*

Interpretation:
"For each TrackId find the total sale quantity bought by the customer with ID = 123."

## Basic Nested (Not Exist Query)

```
SELECT      AL.AlbumId, AL.Title
FROM        Album AL
WHERE   NOT EXISTS
            (SELECT     *
            FROM        Track T, MediaType MT
            WHERE       AL.AlbumId = T.AlbumId
            AND         T.MediaTypeId = MT.MediaTypeId
            AND         MT.Name = 'ACC audio file');
```

The visual diagram for a query is read as follows:

*(1) Select the attributes AlbumId and Title…*   *(2)… from the table Album …*   *(3)… for which there **does not exist any** track whose MediaType name is 'ACC audio file'.*

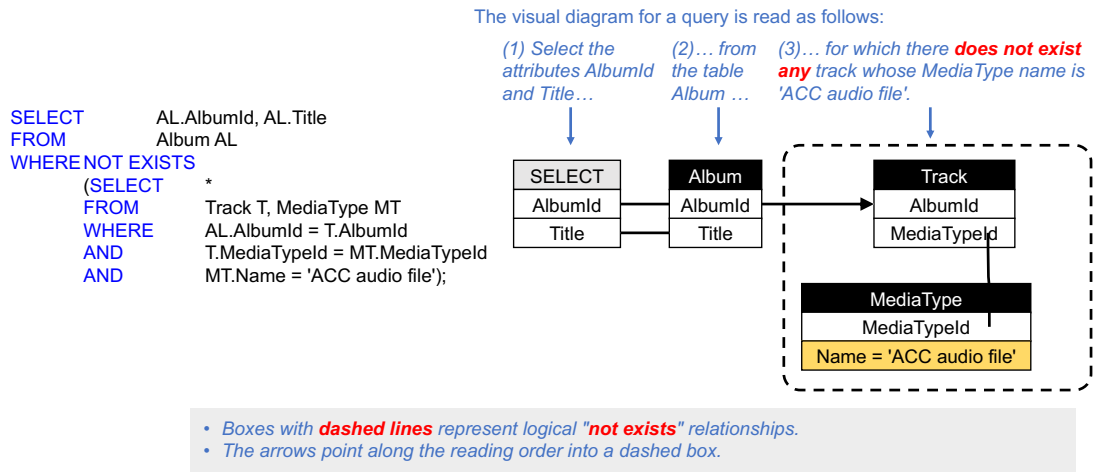

- Boxes with **dashed lines** represent logical "**not exists**" relationships.
- The arrows point along the reading order into a dashed box.

Interpretation:
"Find AlbumId and Title of Albums for which no Track is available as 'ACC audio file' MediaType."

Page: 7/10

## Example double nested SQL query

```
SELECT      A.Name, A.ArtistId
FROM        Artist A
WHERE   NOT EXISTS
            (SELECT     *
            FROM        Album AL
            WHERE       AL.ArtistId = A.ArtistId
            AND     NOT EXISTS
                        (SELECT     *
                        FROM        Track T, MediaType MT
                        WHERE       AL.AlbumId = T.AlbumId
                        AND         T.MediaTypeId = MT.MediaTypeId
                        AND         MT.Name = 'ACC audio file')
            );
```

The visual diagram for a query is read as follows:

*(1) Select the attributes Name and ArtistId …*   *(2)… from the table Artist …*   *(3)… so there **does not exist any** album by those artists …*   *(4) … have **does not have any** track whose MediaType name is 'ACC audio file'*

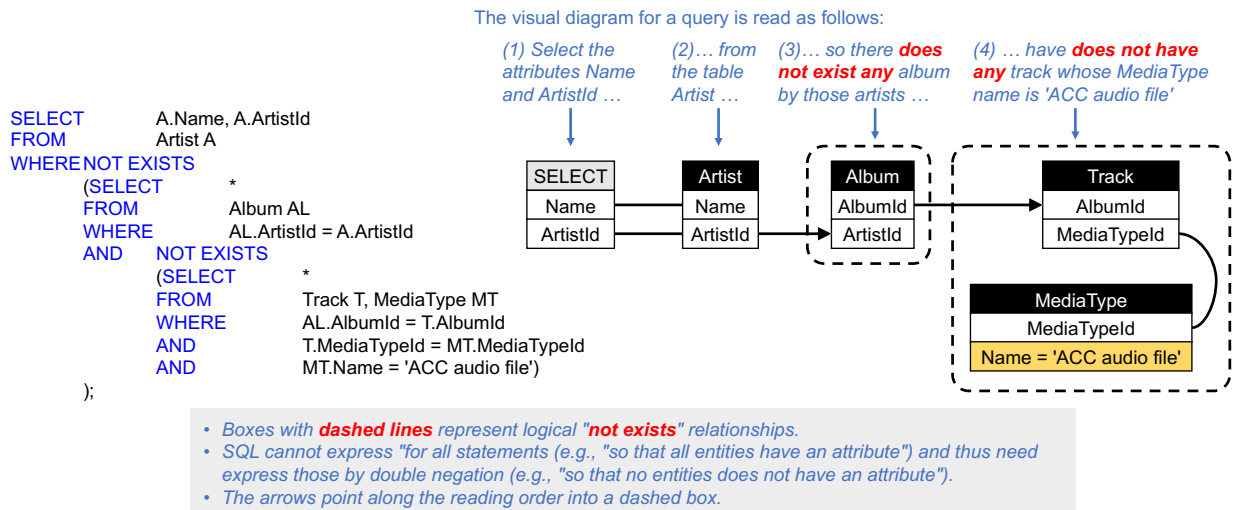

- Boxes with **dashed lines** represent logical "**not exists**" relationships.
- SQL cannot express "for all statements (e.g., "so that all entities have an attribute") and thus need express those by double negation (e.g., "so that no entities does not have an attribute").
- The arrows point along the reading order into a dashed box.

Interpretation:
"Find Name and ArtistId of Artists who have no Album that does not have any Track whose MediaType name is 'ACC audio file'."

Page: 8/10

# Example double nested SQL query ("for all" simplification)

The visual diagram for a query is read as follows:

*(1) Select the attributes Name and ArtistId …*
*(2) … from the table Artist …*
*(3) … for whom **all of** their albums ...*
*(4) … have a track whose MediaType name is 'ACC audio file'*

```
SELECT      A.Name, A.ArtistId
FROM        Artist A
WHERE   NOT EXISTS
        (SELECT      *
         FROM        Album AL
         WHERE       AL.ArtistId = A.ArtistId
         AND     NOT EXISTS
             (SELECT      *
              FROM        Track T, MediaType MT
              WHERE       AL.AlbumId = T.AlbumId
              AND         T.MediaTypeId = MT.MediaTypeId
              AND         MT.Name = 'ACC audio file')
        );
```

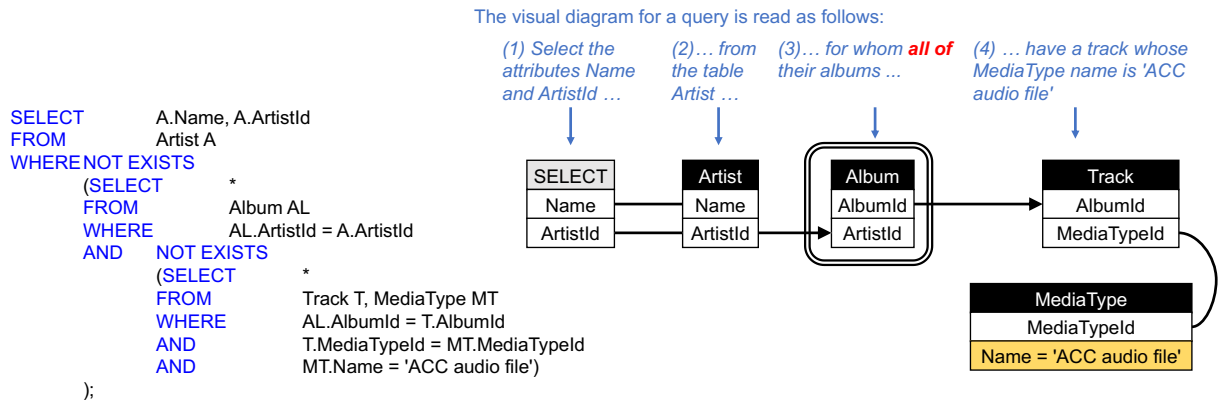

- Boxes with **double lines** represent logical "*for all*" relationships.
- Thus in contrast to SQL, the visual diagrams can express **for all statements** and can thus avoid double negation.
- The arrows point along the reading order into and out of a double lined box.

Interpretation:
"Find Name and ArtistId of Artists for whom all their Albums contain at least one Track whose MediaType name is 'ACC audio file'."

# Legend for visual diagrams

**Joins with an inequality predicate** are shown with a line and label.

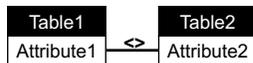

WHERE Table1.Attribute1 <> Table2.Attribute2

Boxes with a yellow background show **selection predicates** on that Attribute and the value being matched.

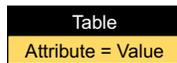

WHERE  Table.Attribute = Value

Boxes with a gray background show **Group By** operations on that Attribute.

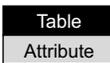

GROUP BY  Table.Attribute

**Logical not exists relations** are shown using a dashed line around the affected tables.

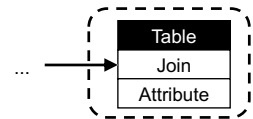

"... so there does not exist any Table with Attribute."

**Logical for all relations** are shown using two solid lines around the affected tables.

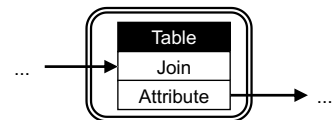

"... so that for all tables with Attribute, ..."

*You can always go back to the tutorial and this summary legend while answering the 12 queries.*

*To do that, click at the Tutorial PDF Link at the bottom banner of the test.*

# F   STUDY DETAILS: 12 TEST QUESTIONS

We include here the 12 test questions that our study participants had to answer. Notice that questions 7–9 include grouping, which is not focus of our paper.

---

## Q1: Conjunctive Query #1

```
SELECT      A.Name
FROM        Artist A, Album AL, Track T
WHERE       AL.AlbumId = T.AlbumId
AND         A.ArtistId = AL.ArtistId
AND         A.Name = T.Composer;
```

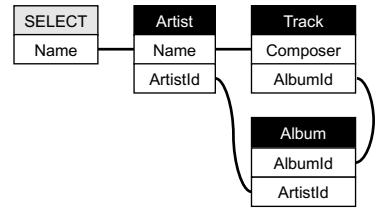

A. Find artists who have an album with a track that is composed by themselves.
B. Find artists who have an album with a track whose composer has the same name as the artists themselves.
C. Find artists whose names are the same as the composer of some track in some album.
D. Find artists whose names are the same as the composer of some track in an album by an artist other than themselves.

---

## Q2: Conjunctive Query #2

```
SELECT      E1.EmployeeId
FROM        Employee E1, Employee E2, Customer C, Invoice I, InvoiceLine IL, Track T, Genre G
WHERE       E1.ReportsTo = E2.EmployeeId
AND         E1.Country <> E2.Country
AND         E2.EmployeeId = C.SupportRepId
AND         I.CustomerId = C.CustomerId
AND         I.InvoiceId = IL.InvoiceId
AND         T.TrackId = IL.TrackId
AND         T.GenreId = G.GenreId
AND         G.Name = 'Rock';
```

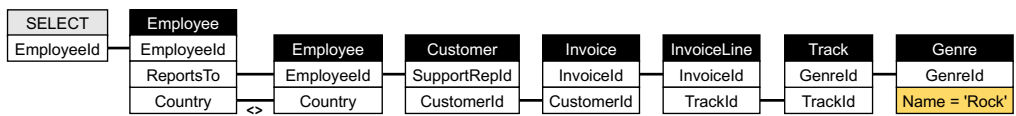

A. Find employees who report to an employee in a different country and the former employee supports at least one customer that has bought a 'Rock' track.
B. Find employees who report to an employee in a different country and the former employee supports only support customers that have bought a 'Rock' track.
C. Find employees who report to an employee in a different country and the latter employee only supports customers that have bought a 'Rock' track.
D. Find employees who report to an employee in a different country and the latter employee supports at least one customer that has bought a 'Rock' track.

## Q3: Conjunctive Query #3

```
SELECT     A.Name
FROM       Artist A, Album AL, Track T,
           PlaylistTrack PT, Playlist P, MediaType MT, Genre G,
           InvoiceLine IL, Invoice I, Customer C
WHERE      AL.ArtistId = A.ArtistId
AND        AL.AlbumId = T.AlbumId
AND        T.TrackId = PT.TrackId
AND        P.PlaylistId = PT.PlaylistId
AND        T.MediaTypeId = MT.MediaTypeId
AND        G.GenreId = T.GenreId
AND        T.TrackId = IL.TrackId
AND        I.InvoiceId = IL.InvoiceId
AND        I.CustomerId = C.CustomerId
AND        MT.Name = 'AAC audio file'
AND        G.Name = 'Rock';
```

A. Find artists who have an album that has a 'Rock' track that is available as 'ACC audio file', and the album has a track that is in a playlist and was purchased by a customer.
B. Find artists who have an album that has a 'Rock' track that is available as 'ACC audio file', is in a playlist, and was purchased by a customer.
C. Find artists who have an album that has a track that is in a playlist and was purchased by a customer, and a 'Rock' track that is available as 'ACC audio file'.
D. Find artists who have an album that has a track that is in a playlist, is available as 'ACC audio file', and was purchased by a customer who also bought a 'Rock' track from the same artist.

## Q4: Self-Join Query #1

```
SELECT     A.ArtistId, A.Name
FROM       Artist A, Album AL1, Album AL2, Track T1, Track T2, Genre G1, Genre G2,
           PlaylistTrack PT1, PlaylistTrack PT2
WHERE      A.ArtistId = AL1.ArtistId
AND        A. ArtistId = AL2. ArtistId
AND        AL1.AlbumId = T1.AlbumId
AND        AL2.AlbumId = T2.AlbumId
AND        T1.GenreId = G1.GenreId
AND        T2.GenreId = G2.GenreId
AND        PT1.PlaylistId = PT2.PlaylistId
AND        PT1.TrackId = T1.TrackId
AND        PT2.TrackId = T2.TrackId
AND        G1.Name = 'Rock'
AND        G2.Name = 'Pop';
```

A. Find artists who have an album with a 'Pop' track and an album with a 'Rock' track and both tracks are in the same playlist.
B. Find artists who have an album with a 'Pop' track and a 'Rock' track and each track is in at least one playlist.
C. Find artists who have an album with a 'Pop' track and an album with a 'Rock' track and each track is in at least one playlist.
D. Find artists who have an album with a 'Pop' track and a 'Rock' track and both tracks are in the same playlist.

## Q5: Self-Join Query #2

SELECT  C.CustomerId, C.FirstName, C.LastName
FROM    Customer C, Invoice I1, Invoice I2
WHERE   C.State = 'Michigan'
AND     C.CustomerId = I1.CustomerId
AND     C.CustomerId = I2.CustomerId
AND     I1.BillingState <> I2.BillingState;

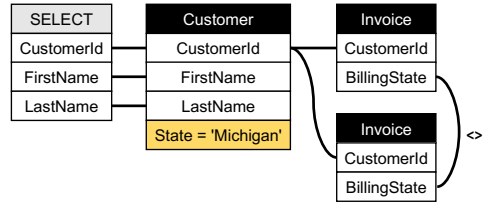

A. Find customers from 'Michigan' that have two invoices billed at two different states where one of them is 'Michigan'.
B. Find customers from 'Michigan' that have two invoices billed at two different states where none of them is 'Michigan'.
C. Find customers from 'Michigan' that have two invoices billed at two different states.
D. Find customers from 'Michigan' that have two invoices billed at 'Michigan'.

## Q6: Self-Join Query #3

SELECT  P.PlaylistId, P.Name
FROM    Playlist P, PlaylistTrack PT1,
        PlaylistTrack PT2, PlaylistTrack PT3,
        Track T1, Track T2, Track T3
WHERE   P.PlaylistId = PT1.PlaylistId
AND     P.PlaylistId = PT2.PlaylistId
AND     P.PlaylistId = PT3.PlaylistId
AND     PT1.TrackId <> PT2.TrackId
AND     PT2.TrackId <> PT3.TrackId
AND     PT1.TrackId <> PT3.TrackId
AND     PT1.TrackId = T1.TrackId
AND     PT2.TrackId = T2.TrackId
AND     PT3.TrackId = T3.TrackID
AND     T1.AlbumId = T2.AlbumId
AND     T2.AlbumId = T3.AlbumId
AND     T2.Composer = T3.Composer;

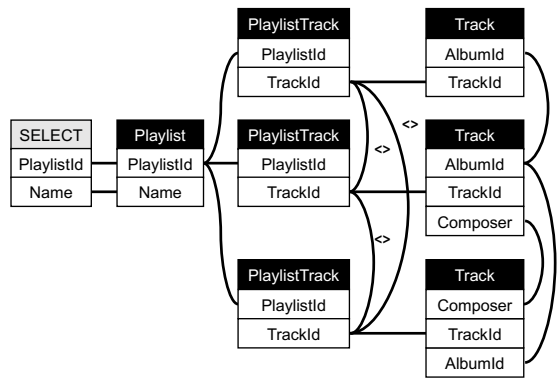

A. Find playlists that have at least 3 different tracks that are in the same album and they are all made by the same composer.
B. Find playlists that have at least 3 different tracks so that at least 2 of them are in the same album but all 3 tracks are made by the same composer.
C. Find playlists that have at least 3 different tracks so that at least 2 of them are in the same album and made by the same composer.
D. Find playlists that have at least 3 different tracks that are in the same album and at least 2 of them are made by the same composer.

## Q7: Grouping Query #1

```
SELECT      I.CustomerId, SUM(IL.Quantity)
FROM        Artist A, Album AL, Track T, InvoiceLine IL, Invoice I
WHERE       A.ArtistId = AL.ArtistId
AND         AL.AlbumId = T.AlbumId
AND         T.TrackId = IL.TrackId
AND         IL.InvoiceId = I.InvoiceId
AND         A.Name = 'Carlos'
GROUP BY    I.CustomerId;
```

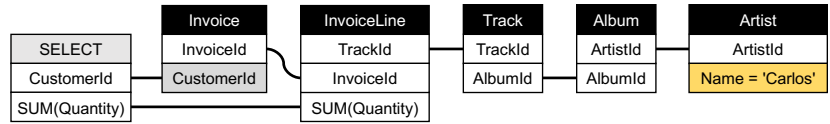

Options
A. For each customer who bought a track from an artist named 'Carlos', find the number of tracks they bought that are by that same artist named 'Carlos'.
B. For each customer who bought a track from an artist named 'Carlos', find the number of tracks they bought that are part of invoices that include a track by that same artist named 'Carlos'.
C. For each customer who bought a track from an artist named 'Carlos', find the total number of tracks that customer has purchased.
D. For each customer who bought a track from an artist named 'Carlos', find the total number of invoices they have.

## Q8: Grouping Query #2

```
SELECT      T.AlbumId, MAX(T.Milliseconds)
FROM        Track T, Playlist P, PlaylistTrack PT, Genre G
WHERE       T.TrackId = PT.TrackId
AND         P.PlaylistId = PT.PlaylistId
AND         T.GenreId = G.GenreId
AND         G.Name = 'Classical'
GROUP BY    T.AlbumId;
```

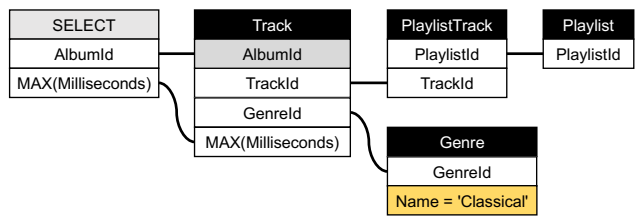

A. For each album that has a 'Classical' track, find the maximum duration of any track that is listed in at least one playlist.
B. For each album that has a 'Classical' track, find the maximum duration of any track that is listed in some playlist that includes a 'Classical' track.
C. For each album that has a 'Classical' track, find the maximum duration of any 'Classical' track that is listed in at least one playlist.
D. For each album that has a 'Classical' track listed in at least one playlist, find the maximum duration of any track in that album.

## Q9: Grouping Query #3

```
SELECT      G.Name, MAX(T.Milliseconds)
FROM        Playlist P, PlaylistTrack PT, Track T, Genre G, InvoiceLine IL, Invoice I, Customer C
WHERE       T.GenreId = G.GenreId
AND         T.TrackId = IL.TrackId
AND         IL.InvoiceId = I.InvoiceId
AND         I.CustomerId = C.CustomerId
AND         PT.TrackId = T.TrackId
AND         P.PlaylistId = PT.PlaylistId
AND         P.Name = 'workout'
AND         C.Country = 'France'
GROUP BY    G.Name;
```

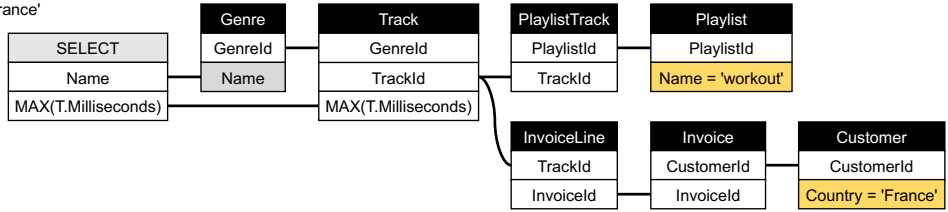

A. For each genre, find the maximum duration of any track that is sold to at least one customer from France who bought some track that is listed in a playlist named 'workout'.
B. For each genre, find the maximum duration of any track that is sold to at least one customer from France and is listed in a playlist named 'workout'.
C. For each genre that has a track listed in a playlist named 'workout', find the maximum duration of any track that is sold to at least one customer from France.
D. For each genre that has a track sold to at least one customer from France, find the maximum duration of any track that is listed in a playlist named 'workout'.

## Q10: Nested Query #1

```
SELECT    A.ArtistId, A.Name
FROM      Artist A
WHERE     NOT EXISTS
          (SELECT *
           FROM    Album AL, Track T
           WHERE   A.ArtistId = AL.ArtistId
           AND     AL.AlbumId = T.AlbumId
           AND     T.Composer = A.Name);
```

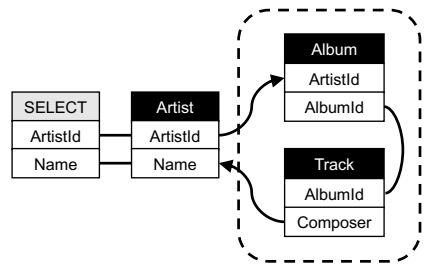

A. Find artists who do not have any album that has a track that is composed by someone with the same name as the artist.
B. Find artists who have an album that does not have any track that is composed by someone with the same name as the artist.
C. Find artists who do not have any album where all its tracks are composed by someone with the same name as the artist.
D. Find artists so that all their albums have a track that is not composed by someone with the same name as the artist.

## Q11: Nested Query #2

```
SELECT     A.ArtistId, A.Name
FROM       Artist A, Album AL1, Album AL2
WHERE      A.ArtistId = AL1.ArtistId
AND        A.ArtistId = AL2.ArtistId
AND        AL1.AlbumId <> AL2.AlbumId
AND        NOT EXISTS
           (SELECT    *
            FROM      Track T1, Genre G1
            WHERE     AL1.AlbumId = T1.AlbumId
            AND       T1.GenreId = G1.GenreId
            AND       G1.Name = 'Rock')
AND        NOT EXISTS
           (SELECT    *
            FROM      Track T2
            WHERE     AL2.AlbumId = T2.AlbumId
            AND       T2.Milliseconds < 270000);
```

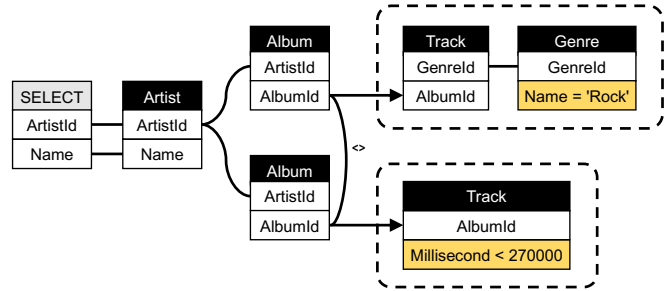

A. Find artists that have at least two albums such that they both do not have any track in the 'Rock' genre and all their tracks are shorter than 270000 milliseconds.
B. Find artists that have at least two albums such that one of their albums does not have any track in the 'Rock' genre and another of their albums only has tracks shorter than 270000 milliseconds.
C. Find artists that have at least two albums such that they both do not have any track in the 'Rock' genre and none of their track is shorter than 270000 milliseconds.
D. Find artists that have at least two albums such that one of their albums does not have any track in the 'Rock' genre and another of their albums does not have any track shorter than 270000 milliseconds.

## Q12: Nested Query #3

```
SELECT     A.ArtistId, A.Name
FROM       Artist A, Album AL
WHERE      A.ArtistId = AL.ArtistId
AND        NOT EXISTS
           (SELECT    *
            FROM      Track T, Genre G
            WHERE     AL.AlbumId = T.AlbumId
            AND       T.GenreId = G.GenreId
            AND       G.Name = 'Jazz'
            AND       NOT EXISTS
                      (SELECT    *
                       FROM      Playlist P, PlaylistTrack PT
                       WHERE     P.PlaylistId = PT.PlaylistId
                       AND       PT.TrackId = T.TrackId)
           );
```

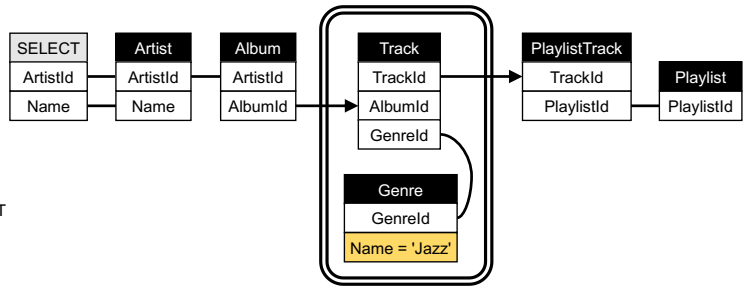

A. Find artists that have an album such that none of its tracks that are in the 'Jazz' genre are individually in at least one playlist.
B. Find artists that have an album such that at least one of its tracks that are in the 'Jazz' genre are in all playlists.
C. Find artists that have an album such that each its tracks that are in the 'Jazz' genre are in all playlists.
D. Find artists that have an album such that each of its tracks that are in the 'Jazz' genre are individually in at least one playlist.

|                              |                                 |                              |
|:----------------------------:|:-------------------------------:|:----------------------------:|
| Sailor (sid,sname,rating,age)| Student (sid, sname)            | Actor (aid, aname)           |
| Reserves (sid, bid, day)     | Takes (sid, cid, semester)      | Plays (aid, mid, role)       |
| Boat (bid, bname, color)     | Course (cid, cname, department) | Movie (mid, mname, director) |
| (a) Sailors                  | (b) Students                    | (c) Actors                   |

Figure 22: Three database schemas: Sailors reserving boats [65], students taking classes, and actors playing in movies.

## G   INTUITIVE EXAMPLES

We bring several examples of visualizations of syntactic SQL variants of similar SQL queries over different schemas. We hope that those examples provide an intuition for the helpfulness of those diagrams for recognizing similar logical patterns of SQL composition.

We use three different database schemas shown in Fig. 22: Sailors reserving certain types of boats, students taking certain types of classes, and actors playing in certain types of movies.

Figure 23 illustrates 3 SQL queries, their logic trees, and their QV. Notice that neither SQL queries, nor the logic trees, allow to quickly distinguish the different queries. In contrast, our digrams emphasize the difference between "no red boat", "all reserved boats need to be red", "all red boats are reserved."

Figure 24 shows 3 syntactically different SQL queries that are equivalent. They are thus equivalent in tuple relational calculus (TRC), have the same logic tree, and the same QV.

Figure 25 applies the logical pattern behind above 3 queries to the two additional schemas. The interpretations on the student database are "students who take no art class", "students who take only art classes", and "students who take all art classes." The interpretations on the actors database are "actors who play in no movie by Hitchcock", "actors who play only in movies by Hitchcock", and "actors who play in all movies by Hitchcock."

Figure 26 shows the same 9 queries as diagrams. Notice the similarity of queries across rows (thus across different schemas) become apparent. This similarity in logical pattern is not as readily visible in SQL.

| | Sailors who do **not** reserve red boats | Sailors who reserve **only** red boats | Sailors who reserve **all** red boats |
|---|---|---|---|
| SQL | ```sql
SELECT S.sname
FROM   Sailor S
WHERE  NOT EXISTS(
   SELECT *
   FROM   Reserves R
   WHERE  R.sid = S.sid
   AND    EXISTS(
      SELECT *
      FROM   Boat B
      WHERE  B.color = 'red'
      AND    R.bid = B.bid))
``` | ```sql
SELECT S.sname
FROM   Sailor S
WHERE  NOT EXISTS(
   SELECT *
   FROM   Reserves R
   WHERE  R.sid = S.sid
   AND    NOT EXISTS(
      SELECT *
      FROM   Boat B
      WHERE  B.color = 'red'
      AND    R.bid = B.bid))
``` | ```sql
SELECT S.sname
FROM   Sailor S
WHERE  NOT EXISTS(
   SELECT *
   FROM   Boat B
   WHERE  B.color = 'red'
   AND    NOT EXISTS(
      SELECT *
      FROM   Reserves R
      WHERE  R.bid = B.bid
      AND    R.sid = S.sid))
``` |
| LT | T: {Sailor S}, P: {}, Selection Attributes: {S.sname} → T: {Reserves R}, P: {(S.sid, =, R.sid)}, Q: $\nexists$ → T: {Boat B}, P: {(B.bid, =, R.bid), (B.color, =, 'red')}, Q: $\exists$ | T: {Sailor S}, P: {}, Selection Attributes: {S.sname} → T: {Reserves R}, P: {(S.sid, =, R.sid)}, Q: $\forall$ → T: {Boat B}, P: {(B.bid, =, R.bid), (B.color, =, 'red')}, Q: $\exists$ | T: {Sailor S}, P: {}, Selection Attributes: {S.sname} → T: {Boat B}, P: {(B.color, =, 'red')}, Q: $\forall$ → T: {Reserves R}, P: {(B.bid, =, R.bid), (S.sid, =, R.sid)}, Q: $\exists$ |
| QV | (QV diagram: SELECT sname ← Sailor[sid, sname] ← Reserves[sid, bid, day] (dashed) ← Boat[bid, bname, color='red']) | (QV diagram: SELECT sname ← Sailor[sid, sname] ← Reserves[sid, bid, day] (double border) ← Boat[bid, bname, color='red']) | (QV diagram: SELECT sname ← Sailor[sid, sname] ← Reserves[sid, bid, day] ← Boat[bid, bname, color='red'] (double border)) |

Figure 23: Sailors reserving no red boats, only red boats, or all red boats.

```sql
SELECT S.sname
FROM   Sailor S
WHERE  NOT EXISTS(
   SELECT *
   FROM   Reserves R
   WHERE  R.sid = S.sid
   AND    NOT EXISTS(
      SELECT *
      FROM   Boat B
      WHERE  B.color = 'red'
      AND    R.bid = B.bid))
```

```sql
SELECT S.sname
FROM   Sailor S
WHERE  S.sid NOT IN(
   SELECT R.sid
   FROM   Reserves R
   WHERE  R.bid NOT IN(
      SELECT B.bid
      FROM   Boat B
      WHERE  B.color = 'red'))
```

```sql
SELECT S.sname
FROM   Sailor S
WHERE  NOT S.sid = ANY(
   SELECT R.sid
   FROM   Reserves R
   WHERE  NOT R.bid = ANY(
      SELECT B.bid
      FROM   Boat B
      WHERE  B.color = 'red'))
```

Figure 24: Three semantically equivalent yet syntactic different variants for "sailors who reserve only red boats." All three SQL queries lead to the same logic tree (LT) and the same QV.



|  | not | only | all |
|---|---|---|---|
| Sailors | ```
SELECT S.sname
FROM   Sailor S
WHERE  NOT EXISTS(
   SELECT *
   FROM   Reserves R
   WHERE  R.sid = S.sid
   AND    EXISTS(
      SELECT *
      FROM   Boat B
      WHERE  B.color = 'red'
      AND    R.bid = B.bid))
``` | ```
SELECT S.sname
FROM   Sailor S
WHERE  NOT EXISTS(
   SELECT *
   FROM   Reserves R
   WHERE  R.sid = S.sid
   AND    NOT EXISTS(
      SELECT *
      FROM   Boat B
      WHERE  B.color = 'red'
      AND    R.bid = B.bid))
``` | ```
SELECT S.sname
FROM   Sailor S
WHERE  NOT EXISTS(
   SELECT *
   FROM   Boat B
   WHERE  B.color = 'red'
   AND    NOT EXISTS(
      SELECT *
      FROM   Reserves R
      WHERE  R.bid = B.bid
      AND    R.sid = S.sid))
``` |
| Students | ```
SELECT S.sname
FROM   Student S
WHERE  NOT EXISTS(
   SELECT *
   FROM   Takes T
   WHERE  T.sid = S.sid
   AND    EXISTS(
      SELECT *
      FROM   Class C
      WHERE  C.department = 'art'
      AND    C.cid = T.cid))
``` | ```
SELECT S.sname
FROM   Student S
WHERE  NOT EXISTS(
   SELECT *
   FROM   Takes T
   WHERE  T.sid = S.sid
   AND    NOT EXISTS(
      SELECT *
      FROM   Class C
      WHERE  C.department = 'art'
      AND    C.cid = T.cid))
``` | ```
SELECT S.sname
FROM   Student S
WHERE  NOT EXISTS(
   SELECT *
   FROM   Class C
   WHERE  C.department = 'art'
   AND    NOT EXISTS(
      SELECT *
      FROM   Takes T
      WHERE  T.cid = C.cid
      AND    T.sid = S.sid))
``` |
| Actors | ```
SELECT A.aname
FROM   Actor A
WHERE  NOT EXISTS(
   SELECT *
   FROM   Casts C
   WHERE  C.aid = A.aid
   AND    EXISTS(
      SELECT *
      FROM   Movie M
      WHERE  M.director = 'Hitchcock'
      AND    M.mid = C.mid))
``` | ```
SELECT A.aname
FROM   Actor A
WHERE  NOT EXISTS(
   SELECT *
   FROM   Casts C
   WHERE  C.aid = A.aid
   AND    NOT EXISTS(
      SELECT *
      FROM   Movie M
      WHERE  M.director = 'Hitchcock'
      AND    M.mid = C.mid))
``` | ```
SELECT A.aname
FROM   Actor A
WHERE  NOT EXISTS(
   SELECT *
   FROM   Movie M
   WHERE  M.director = 'Hitchcock'
   AND    NOT EXISTS(
      SELECT *
      FROM   Casts C
      WHERE  C.mid = M.mid
      AND    C.aid = A.aid))
``` |

Figure 25: The interpretation of the 3 SQL queries from Fig. 23 applied to 2 additional schemas.

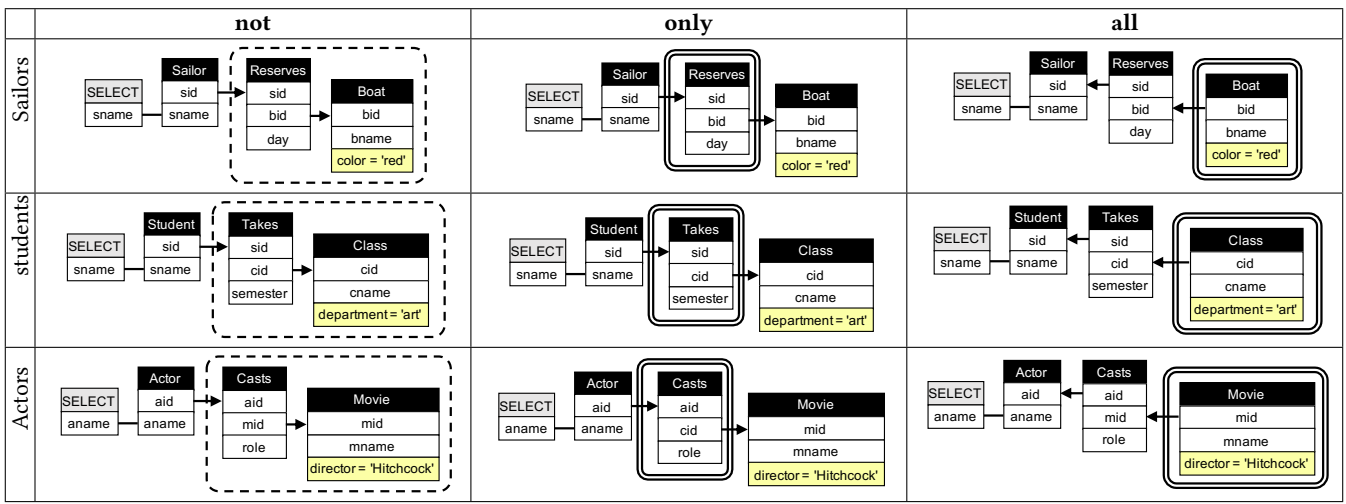

Figure 26: The 9 SQL queries from Fig. 25 repeated as diagrams. Notice that the similarity of the *SQL pattern* becomes visible across rows and thus across different schemas. This similarity is not as readily apparent in SQL only.